\documentclass[11pt, a4paper]{article}

\usepackage[T1]{fontenc}
\usepackage[utf8]{inputenc}
\usepackage[backend=bibtex]{biblatex}   
\usepackage[english]{babel}
\usepackage{appendix}
\usepackage{graphicx}
\usepackage{geometry}
\usepackage{qcircuit}
\usepackage{amssymb}
\usepackage{amsthm}
\usepackage{amsmath}
\usepackage{mathtools}
\usepackage[colorlinks]{hyperref}
\usepackage[nameinlink,noabbrev]{cleveref}
\usepackage{float}
\usepackage[notrig]{physics}
\usepackage{color}
\usepackage{tikz}
\usepackage{xy}
\usepackage{fancybox}
\usepackage{float}
\usepackage{mathtools}
\usepackage[format=plain,labelfont={bf,it},textfont=it]{caption}
\usepackage{subcaption}
\usepackage{algorithm,algpseudocode}
\usepackage{algorithmicx}
\usepackage{diagbox,multirow}
\usepackage{enumerate}
\newcommand\clearrow{\global\let\rowmac\relax}
\usepackage{tikz}
\newcommand{\circled}[1]{\raisebox{.3pt}{\textcircled{\raisebox{-.8pt}{\footnotesize #1}}}}

\usetikzlibrary{positioning,arrows,decorations.pathmorphing,shapes,patterns}
\usetikzlibrary{decorations.markings}

\tikzstyle{abo}=[rectangle,rounded corners,inner sep=.25em, text centered, text width=.7em, draw=gray,font=\small]
\tikzstyle{bb}=[dashed,rectangle,anchor=north west,text centered,minimum
width=4.1cm,minimum height=3.25cm,draw=black]
\tikzstyle{bbk}=[dotted,rectangle,text centered,anchor=north west,minimum width=1.8cm,minimum height=1.4cm,draw=black]
\tikzstyle{bbkd}=[dotted,rectangle,text centered,minimum width=2.15cm,minimum height=1.65cm,draw=black]

\tikzset{split fill/.style args={#1 and #2}{path picture={
    \fill [#1] (path picture bounding box.south west)
      rectangle (path picture bounding box.north);
    \fill [#2] (path picture bounding box.south)
      rectangle (path picture bounding box.north east);
}}}

\addto\extrasenglish{    }

\floatstyle{ruled}
\newfloat{protocol}{htbp}{lok}
\floatname{protocol}{Protocol}
\newfloat{algorithm}{htbp}{lok}
\floatname{algorithm}{Algorithm}
\newfloat{subroutine}{htbp}{lok}
\floatname{subroutine}{Subroutine}

\crefname{algorithm}{Algorithm}{Algorithms}
\crefname{subroutine}{Subroutine}{Subroutines}
\crefname{protocol}{Protocol}{Protocols}
\crefname{definition}{Definition}{Definitions}
\crefname{theorem}{Theorem}{Theorems}
\crefname{corollary}{Corollary}{Corollary}

\usepackage{tabularx,ragged2e,booktabs}
\newcolumntype{C}[1]{>{\centering\arraybackslash}m{#1}}
\newcolumntype{C}[1]{>{\centering\arraybackslash\hspace{0pt}}p{#1}}

\DeclarePairedDelimiter\floor{\lfloor}{\rfloor}

\newcommand{\eg}{\textit{e}.\textit{g}.,~}

\tikzset{
  vertex/.style={fill,draw,inner sep=0pt,label distance=1pt,minimum
  size=4pt,circle},
   clusterm/.style={
     column sep=.7cm, row sep=.7cm,
     matrix of nodes,
     nodes=vertex}
}
\newsavebox{\circbox}
\usetikzlibrary{fit,calc,positioning,decorations.pathreplacing,matrix}
\newcommand*{\tightdisplaymath}{\abovedisplayskip\z@\belowdisplayskip\z@}

\newcommand{\boxsize}{.3cm}
\newcommand{\boxabc}[1]%
  {\ovalbox{\text{\begin{minipage}{\boxsize}\centering #1\end{minipage}}}}

\title{Three-qubit exact Grover within the blind oracular quantum computation scheme}
\author{Cica Gustiani, David P. DiVincenzo}
\date{\today}

\begin{document}
\maketitle

\begin{abstract}
Here we extend the concept of blind client-server quantum computation,
in which a client with limited quantum power controls the execution of a
quantum computation on a powerful server, without revealing any details
of the computation.  Our extension is to introduce a three-node setting
in which an oracular quantum computation can be executed blindly.  In
this Blind Oracular Quantum Computation (BOQC), the oracle (Oscar) is
another node, with limited power, who acts in cooperation with the
client (Alice) to supply quantum information to the server so that the
oracle part of the quantum computation can also be executed blindly. We
develop tests of this protocol using two- and three-qubit versions of
the exact Grover algorithm (i.e., with database sizes $4\leq N\leq 8$),
obtaining optimal implementations of these algorithms within a gate
array scheme and the blinded cluster-state scheme.  We discuss the
feasibility of executing these protocols in state-of-the-art three-node
experiments using NV-diamond electronic and nuclear qubits.
\end{abstract}

\section{Introduction}

While the promise of distributed quantum information processing was
already foreseen in theoretical work many decades
ago~\cite{bennett1984quantum, wiesner, cleve1999quantum}, we have
finally entered a time when some of these ideas can be implemented in
the laboratory \cite{hensen2015loophole}.  With these developments, it
is timely to look at the theoretical situation in a new light, and to
evaluate what can be done with the currently very limited resources that
are available.   

In this paper we lay out a concrete plan for putting several new
concepts in distributed quantum computing into action.  Blind quantum
computation~\cite{broadbent2009universal} is an example of a protocol in
which quantum physics gives unique security properties in a distributed
computing setting.  It is a client-server scheme, in which a client with
limited computing power wishes to make use of a powerful server, but in
such a way as to assure that the server is ``blind'', i.e., not able to
determine what computation the client is running, and not able to come
into possession of any intelligible input or output data for this
computation.  It has been shown that an adaptation of the technique of
cluster-state quantum
computation~\cite{raussendorf2000quantum,raussendorf2003measurement,nielsen2006cluster}
can achieve client-server blind quantum computation, and one aspect of
our work in this paper will be to lay out the possibilities for
achieving this in a distributed quantum device involving diamond NV
centers.

It has been standard for twenty years to use oracular algorithms as test
cases for quantum computing implementations.  We will adopt this
approach here as well, but we propose, here and in a companion detailed
paper~\cite{boc}, a new approach to integrating oracular computations into the
client-server paradigm.  In a distributed setting, it is meaningful to
consider the oracle to be a distinct node of a network.  In the case of
the Grover quantum computation, this means a node in possession of an
actual physical database.  Thus we propose here, and explore for
implementation, a three-party distributed computation setting: the
client (Alice), who wants to know the answer to a database-lookup
problem; the oracle (Oscar) who is in possession of this database, and
is willing to reveal information about it to a server, but in a blinded
fashion that will only be intelligible to Alice; and finally the server
itself (Bob), in possession of a powerful quantum computer, with the
capacity to receive remote qubits from Alice and Oscar, to perform
entangling operations, and to broadcast the results of quantum
measurements, under instructions from Alice and Oscar.

Of course, many experiments have achieved some implementation of the
two-qubit Grover algorithm. For instance,
\cite{chuang1998experimental,jones1998implementation,anwar2004implementing}
used the NMR technique, \cite{brickman2005implementation,feng2001grover}
used trapped ions, \cite{dicarlo2009demonstration} used superconducting
qubits, and~\cite{yao2011scheme} used Abelian anyons (in a simulation).
Moreover, \cite{walther2005experimental,chen2007experimental}
demonstrated the algorithm with the one-way quantum
computer~\cite{raussendorf2000quantum,raussendorf2003measurement},
computation scheme~\cite{broadbent2009universal} was demonstrated
in~\cite{barz2012demonstration} with just four photonic qubits.  But we
believe that current developments in the quantum technology of
distributed processing using remote NV
centers~\cite{humphreys2018deterministic,hensen2015loophole} make our
three-party version of blind client-server quantum computation feasible
for a full implementation study.

We can indicate precisely how this implementation can be achieved for
the standard two-qubit Grover problem (and will do so in the final
section), but we will primarily use the present study to analyse the
implementation of scaled-up oracle problems.  Thus, we will examine in
detail the possible realizations of three-qubit Grover.  It is already
known that going from two to three qubits adds challenges for the
implementation: two calls to the oracle are needed rather than one.  In
addition, the original Grover procedure does not give an error-free
identification of the database state, except in the single case of the
two-qubit case~\cite{dicarlo2009demonstration}. This problem was solved
by subsequent modifications of Grover's
procedure~\cite{chi1999quantum,hoyer2000arbitrary,long2001grover,liu2014exact},
and we take account in the present work of these modifications needed to
make the database search an ``exact'' algorithm. 

Given the various inconvenient features of three-qubit Grover --- two
oracle calls, lack of exactness, necessity for two-qubit gates at all
stages of the algorithm ---  it is not surprising that there has been only a limited set of attempts to implement in the laboratory, and never in a distributed or blind setting.  But \cite{figgatt2017complete} represented quantum circuits for different number of queries, \cite{yang2007implementation} illustrated implementation with cavity quantum electrodynamics, an experiment using NMR was performed in~\cite{vandersypen2000implementation}, and~\cite{figgatt2017complete} demonstrated using trapped atomic ions~\cite{debnath2016demonstration}.  However, to maintain the certainty in going from two-bit to three-qubit Grover, more complex gates are required.  Only one experiment so far demonstrated the three-qubit exact Grover, which used a magnetic resonance system~\cite{liu2015first}.  
 
But as we show below, ``three-qubit Grover" in fact encompasses a very large set of potential algorithms, and we explore these possibilities systematically here, with the objective of identifying the easiest implementations in the NV-center setting.  The multiplicities of these Grover algorithms come in several forms.  First, the number of database entries can be as many as $N=2^3=8$, but it can be fewer.  Each of the new cases N=5, 6, 7, and 8, is a separate problem, and we consider all of these here.  While for $N=8$ all of the three-qubit states are in use, for $N<8$ only a subset are used; the exact choice of this subset is another variable that we have studied one by one. 

There is a final variation of the algorithm that, to our knowledge, has
not been exploited before.  It is not necessary that the number of
distinct entries in the database of Oscar be equal to the number of
entries used in the quantum register.  For example, suppose that Oscar
has five database entries, A, B, C, D, and E.  He and Alice may agree on
an encoding in which A can correspond to marking either the three-qubit
memory location 000 or 001, while the other four have a unique location,
say B $\rightarrow$ 010, C $\rightarrow$ 011, D $\rightarrow$ 100, and E
$\rightarrow$ 101. Then, if Alice's final measurement reveals either 000
or 001, she infers that datum A is stored in Oscar's database.  The
algorithm will also be successful even if Alice cannot reliably
distinguish between the 000 and 001 outcomes, so long as they are
reliably distinguished from the others.  For this reason, we refer to
this approach below as the ``POVM strategy''.  

We have also exhaustively optimized over possible POVM strategies.  We
find that the most economical three-qubit Grover algorithm to implement
is in fact exactly the one that we have just given as an example!  It is
perhaps surprising that using N=6 with only five data is preferable to
simply using N=5, but we find that the POVM freedom allows for reduction
of the gate complexity of the implementation, and thus in the cluster
state implementation.

An unfortunate message is that even this most economical case among all
the three-qubit Grover algorithms is still much more resource intensive
than the two-qubit Grover algorithm.  This increase is modest in the
number of physical qubits used (4 vs. 3), but very large in the number
of gate operations and repeated re-use of physical qubits (approx. 10x
more), and correspondingly large in its coherence demands.  Thus, it
appears that within the Grover family of algorithms, a large jump in the
implementation is unavoidable.  To make these jumps smaller, it will be
necessary to look at other families of oracle algorithms.

\section{Preliminaries}
\subsection{One-way quantum computer and universal blind quantum
computation}
\label{sec21}

In this section two measurement-based quantum computation schemes ---
some of our works are based on these schemes --- are recalled: the
one-way quantum computer (1WQC)~\cite{raussendorf2000quantum} and the
universal blind quantum computation
(UBQC)~\cite{broadbent2009universal}.  In principle, UBQC allows a
client with small quantum power to delegate her private computation to
an untrustworthy server; the server is a one-way quantum computer,
namely a cluster state computer. 

By contrast to conventional quantum computation, viz. the gate model, a
computation within the 1WQC scheme is performed by \emph{adaptively}
measuring a cluster state.  Adaptive means that the measurement basis
can be dependent on previous measurement outcomes. Therefore, in the
1WQC scheme, a \emph{cluster state} defines the quantum computer, and
consecutive measurements define quantum operations.  A cluster state is
represented as an open graph $\mathcal G$, together with a set of input
nodes  $I$ and a set of output nodes $O$, where $I$ and $O$ may
intersect.  The non-input nodes can be interpreted as qubits whose
states are set in the $xy$-plane of the Bloch sphere; the initialization
of the input nodes are determined by the quantum algorithm.  The edges
of the graph $\mathcal G$ correspond to CPHASE gates operated on the
corresponding node qubits.  The measurements are parameterized with
angles $\vec\phi$; the measurement operators are in the form
$\{\ketbra*{+_{\phi}},\ketbra*{-_{\phi}}\}$, where $ \label{eq:pmtheta}
\ket*{\pm_{\phi}}\coloneqq \frac{\ket0\pm e^{i\phi}\ket1}{\sqrt2},$ and
$\phi\in[0,2\pi)$.  From this point on, we refer to ``measure in angle
$\phi$'' as a projective measurement in basis $\ket*{\pm_\phi}$.
Henceforth, to represent such a computation, we express it as a set
$\{(\mathcal G, I,O),\vec\phi\}$. 

Since quantum measurements unavoidably introduce indeterminacy, adaptive
measurements are performed to obtain deterministic quantum operations. A
measurement angle $\phi_j$ can be $X$- or $Z$-dependent on outcome $i$,
which means correcting $\phi_j$ to $(-1)^{s_i}\phi_j$ or $\phi_j+s_i\pi$
respectively. Here $s_i\in\{0,1\}$ is the outcome of measurement $i$.
This correcting scheme is nicely captured by the notion of
\emph{flow}~\cite{danos2006determinism}, that is a map $f:I^{c}\mapsto
O^{c}$ following certain criteria ($A^c$ means the complement of set
$A$). Thus, measuring $j$, $f(j)$ determines $X$
correction and neighbors of $f(j)$ determine $Z$ corrections. 

Now Alice as a client wants to run her private quantum computation on
the untrusted server of Bob, thus they run the UBQC protocol as follows.
First, Alice has her computation in mind $\{(\mathcal G, I,
O),\vec\phi\}$; she informs Bob only the graph's form. She transmits her
input qubits to Bob then transmits the rest of the qubits, which are the
non-input nodes in $\mathcal G$, in the state $\{\ket*{+_{\theta_j}}\}$,
$j\in I^c$, where $\theta_j\in[0,2\pi)$ is randomly generated from a
discrete set. Second, Bob entangles the received qubits according to the
edges of graph $\mathcal G$ by applying CPHASE gates. Third, Bob
measures every node in $\mathcal G$ in angle $\delta_j$ that is publicly
announced by Alice, where $\delta_j\coloneqq \phi_j+\theta_j+r_j\pi$,
where $r_j\in\{0,1\}$ is randomly generated.  Bob announces every
outcome $b_j$ after measuring node $j$.   

The key feature contributing to the blindness of UBQC is the randomness
introduced into several of the variables: $\vec\theta$ which hides the
measurement angles, and $\vec r$ which hides the measurement outcomes.
Since $\vec\theta$ is a vector of parameters describing a set of
non-orthogonal quantum states, inferring $\vec\theta$ is impossible
without disturbing the quantum states. Since $\vec r$ is randomly
generated and is independent of $\vec\delta$, knowing an actual
measurement outcome $s_j\coloneqq b_j\oplus r_j$ from $\vec\delta$ and
$\vec b$ is impossible. Thus, no information is gained by Bob during the
protocol run without disturbing the quantum states.  

\subsection{The exact Grover-H\o yer search algorithm}
\label{sec:exact_grover-hoyer}

The optimality of the Grover algorithm is well known~\cite{zalka1999grover};
high success probability is achieved with the fewest iterations.  As the
number of items in the database $N$ increases, the success probability approaches one,
whereas for small $N$ the error is appreciable.  For instance,
success probabilities ($p_N$) running 3-qubit Grover are: $p_5=0.968$
with 1 iteration,  $p_6=0.907$ with 1 iteration, $p_7=0.871$ with 2
iterations, and $p_8=0.945$ with 2 iterations.  Because of this problem, many workers
devised  modifications or generalizations of the Grover algorithm to achieve
probability one.  For instance, H\o yer \cite{hoyer2000arbitrary}
introduced arbitrary phase rotation in quantum amplitude
amplification, Chi and Kim\cite{chi1999quantum} introduced the single query
search for the case when one quarter of the database is marked,
Long\cite{long2001grover} improved Chi and Kim's algorithm using a phase
matching condition that works for databases with size $2^n$,
followed by Liu\cite{liu2014exact} who generalized it for an arbitrary
size and combination of databases. This section provides details of the
so-called \emph{Grover-H\o yer algorithm}, which combines previous Grover and H\o yer
procedures to achieve probability one --- later we develop a new
algorithm based on that, which also features oracle separation, blindness,
and measurement freedom.
 
Suppose $n$ qubits are used to represent all indices
$x=\{0,\dots,2^{n-1}\}$.  One may arbitrarily choose $N$ elements of $x$
that represent indices of a database $w$, thus $w\subset x$, where
$\abs*{w}=N$, and we will consider the case $2^{n-1}< N \leq 2^n$.  Without loss of generality, we
start from a product of zero states $\ket{0}^{\otimes n}$.  We consider an
operator $A$ that maps a product state into an equal superposition of
$N$ states, thus $A \ket{0}^{\otimes n}=(1/\sqrt{N})/\sum_{j\in
w}\ket{j}\eqqcolon\ket*{\Psi_{in}}$. Suppose we have marked items in the
database $\tau\subset w$ --- we are interested in a special case where
$\abs*{\tau}=1$, thus $\tau\in w$.  Given an oracle that evaluates a
function $f(j)$ that indicates if $j$ indexes a marked item of database,
$f$ induces a partition in the Hilbert space into ``solutions''
($\tau$) and ``non-solutions'' ($w\setminus \tau$) subspaces.  Rewrite
the state $\ket{\Psi_{in}}=\sqrt a
\ket*{\tilde\Psi_1}+\sqrt{1-a}\ket*{\tilde\Psi_0}$, where $a=1/N$, where
$\ket*{\tilde\Psi_1}$ and $\ket*{\tilde\Psi_0}$ are the normalized
states corresponding to $\ket{\Psi_1}\coloneqq\sum_{j\in\tau}\ket j$ and
$\ket{\Psi_0}\coloneqq\sum_{j\in w/\tau}\ket j$.  Henceforth, we will work in the
Hilbert space defined as the subspace spanned by basis
$\{\ket*{\tilde\Psi_0},\ket*{\tilde\Psi_1}\}$.

\begin{algorithm}[!bt]
\caption{Grover-H\o yer algorithm}
\label{alg:exactgrover}
\begin{algorithmic}[1]
    \Require $w,\tau$ 
    \Statex{\bfseries (1) Classical processing}
    \State $N\leftarrow$ size of $w$
    \Ensure  $2^{n-1}< N \leq 2^n$ and $\tau\in w$
    \State $\theta_0 \leftarrow\arcsin{(1/\sqrt{N})}$
    \State $m\leftarrow\floor{(\frac{\pi}{2}-\theta_0)/2\theta_0}$ number of Grover runs
    \State $\theta\leftarrow\frac{\pi}{2}-2m\theta_0$ the remaining rotation
    \State $ \psi, \varphi, u \leftarrow$ \autoref{eq:psi}, \autoref{eq:varphi}, \autoref{eq:u}. 
    \State {$A\gets\ket*{\Psi_{in}}(\bra{0})^{\otimes n}$ }
    \State \label{line:d}{$\{D(\pi),D(\psi)\} \gets$ \autoref{eq:d}}
    \State {$\{O(\pi),O(\varphi+u)\}\gets$ \autoref{eq:o}}
    \State{Obtain a set of necessary operators $\mathcal B$
      (\autoref{eq:b}), which are expressed within operations
      that can be done with the corresponding quantum computer. }\label{line:transform} 
    \Statex{\bfseries (2) Quantum processing}
    \State $\ket{\Psi}\leftarrow {\mathcal A}\ket{0}^{\otimes n}$ 
    \For{$j=1$ to $m$}
    \State $\ket{\Psi}\leftarrow \mathcal D(\pi)\mathcal O(\pi)\ket{\Psi}$,
    \EndFor
    \State \label{line:phiu}$\ket{\Psi}\leftarrow \mathcal D(\psi)\mathcal O(\varphi+u)\ket{\Psi}$
    \State Measure $\ket\Psi$
    \State{\bfseries Exit}
  \end{algorithmic}
\end{algorithm}

Using previously described variables, running~\Cref{alg:exactgrover}
within database $w$ will reveal the marked item  $\tau$ with probability
one.  The main idea of the algorithm is to combine the Grover algorithm
with H\o yer's arbitrary phase rotation (also known as H\o yer amplitude
amplification), which performs the necessary rotation to bring the state
vector exactly into the solution space.  The modified iteration
introduces new operators $\{O(\varphi+u),D(\psi)\}$, where \begin{align}
O(\varphi)&=-I+(1-e^{i\varphi})\ketbra{\tau}\label{eq:o}\\
D(\psi)&=-I+(1-e^{i\psi})\ketbra{\Psi_{in}}\label{eq:d}.  \end{align}
The algorithm comprises two stages: \textbf{classical processing}, where
compatible set of operations for every required unitary is obtained:
\begin{equation}\{\mathcal A, \mathcal O(\pi), \mathcal D(\pi),\mathcal
  O(\varphi+u),\mathcal D(\psi)\}\eqqcolon \mathcal B, \label{eq:b}
\end{equation} and \textbf{quantum processing}, where the quantum
computation is performed on the quantum computer;  every operator in
$\mathcal B$ respectively correspond to unitary matrices in
$\{A, O(\pi),$ $D(\pi), O(\varphi+u)\}$. When we say that we have a
compatible set of operations $\mathcal A$ corresponding to the unitary
operator $A$ (and similarly for all elements of $\mathcal B$), we mean
that we specify an explicit implementation of $A$ as a sequence of
operations $\mathcal A$ that can be performed for some model of quantum
computation, \eg in the form of quantum gates or operations on a cluster
state.  For instance, our result in~\autoref{fig:circuit6} works on a
quantum computer which performs CNOT and arbitrary 1-qubit gates.

The H\o yer amplitude amplification is described by an operator
$Q(\varphi,\psi)=D(\psi)O(\varphi)$, which rotates a state closer to the
solution space by as much as $\theta$, where $\abs{\sin(\theta)}\leq
\sin(2\theta_0)$, $\theta_0=\arcsin(1/\sqrt{N})$. H\o yer found $\varphi$ and $\psi$ such that $Q(\varphi,\psi)$ performs the
desired rotation:
\begin{align}
  \psi&=\arccos\left(1-\frac{\sin^2(\theta)}{2a(1-a)}\right)\label{eq:psi}\\
  \varphi&=2\arctan(\psi/2)(1-2a)\label{eq:varphi} \\
  u &=\arg\left(-a(1-e^{i\psi})-e^{i\psi}\right)-\arg\left(
  (1-e^{i\psi})\sqrt{a(1-a)} \right)\label{eq:u}.
\end{align}
Using those angles, $Q(\varphi,\psi)$ rotates the state
by angle $\theta$ up to some phases $\pm u$;
\begin{equation}
Q(\varphi,\psi)=
\begin{pmatrix}
1 & 0 \\
0 & e^{iu}
\end{pmatrix}
\begin{pmatrix}
\cos\theta & -\sin\theta \\
\sin\theta & \cos\theta 
\end{pmatrix}
\begin{pmatrix}
1 & 0 \\
0 & e^{-iu}
\end{pmatrix}. \label{eq:q}\\
\end{equation}

The unwanted phases $\pm u$ can be cancelled by performing the sequence
$P(-u)Q(\varphi,\psi)P(u)$, where $P(\alpha)=
-I+(1-e^{i\alpha})\ketbra*{\tau}$.  Since the form of operator $P$ is identical to that of $O$, 
$O(\varphi)P(u)=O(\varphi + u)$ (see step~\ref{line:phiu} of ~\cref{alg:exactgrover}). When H\o yer amplitude amplification is
applied in the last iteration of the Grover algorithm, the state is entirely aligned to the
solution space after the application of $P(-u)Q(\varphi,\psi)$. Thus,
applying $P(u)$ afterward will change only the global phase of the
state. This is the reason for omitting the last phase correction
in~\cref{alg:exactgrover}.  

\subsection{Exhaustive search for most economical Grover algorithm}
\label{sub:exhaustive}

The challenge in realizing the Grover-H\o yer algorithm --- apart from
running the quantum processing with arbitrarily small error --- is the
optimization of the circuit preparation indicated on
line~\ref{line:transform} of~\cref{alg:exactgrover}, where the desired
unitary map must be written out as a set of quantum gates that can be
run in the quantum computer.  We develop an approach based on DiVincenzo
and Smolin~\cite{divincenzo1994results} (DS94) to overcome this
challenge --- such a challenge will appear again later when we need to
obtain a graph state. This section mainly reviews DS94.

DS94 is a systematic, exhaustive approach: given the desired unitary
map $M$, where $M\in$ SU(8), a set of 2-qubit gates networks are
optimized over, where every 2-qubit gate is in SU(4). We refer to
``topology'' of a 2-qubit gate network as a configuration of those 2-qubit
gates.   As we are concerned here with a 3-qubit
operations, as was also the case in the study of DS94, the notations of DS94 are used:
qubits are indicated with numbers 1,2, and 3; a 2-qubit gate is indicated
with the number of the untouched qubit. A topology is denoted by numbers within
parenthesis, where each number represents the corresponding 2-qubit gate.  So, for example,
topology $(321)$ indicates 2-qubit gates applied on qubits: $\{1,2\}$,
$\{1,3\}$, and $\{2,3\}$; note that the order of gates here is relevant, since these gates do not commute.   

To efficiently obtain an exhaustive set of topologies, all possible topologies of 2-qubit
gate networks are
enumerated, then the equivalent ones are eliminated. Two different
topologies can be equivalent for the following
reasons\cite{divincenzo1994results}: \emph{time-reversal} which means
placing the gates in time-reversed order, \eg 
$(12123)=(32121)$; \emph{bit-relabelling} \eg relabeling qubit 1 and 2,
thus $(12123)=(21213)$; and \emph{conjugation by swapping}
which means swapping of the states of any pair of bits, \eg
$(12123)=(13123)=(12323)=(12313)$.  For the systems with an unused subspace
in the Hilbert space --- thus for $N<8$, where $N$ is the
dimension of the Hilbert space --- the reordering must preserve the
state space.  For instance, database $w=\{0,1,2,4,7\}$ is conserved with
permutation of every element in $S_3$ --- this is easiest seen by writing 
this set $w$ in three-bit notation, $w=\{000,001,010,100,111\}$.  On the other hand, database
$w=\{0,1,2,3,4\}$ is conserved only with one permutation of $S_3$:
$\begin{psmallmatrix*}1&2&3\\1&3&2\end{psmallmatrix*}$.

The non-linear minimization Broyden-Fletcher-Goldfarb-Shanno
(BFGS)~\cite{press1986numerical} is used for the optimization in DS94
with the objective function defined as
$f=\sum_{i}\sum_{j}\abs{M_{ij}-S_{ij}}^2, $ where $M$ is the desired
SU(8) unitary, and $S$ is the matrix resulting from composing the
2-qubit gate network. The minimization is over the parameters of the
individual SU(4) matrices describing the two-qubit gates.  It is
successful if $f=0$ to a reasonable accuracy; thus a 2-qubit gate
network that implements $M$ is found.

\section{Results}
\subsection{The blind oracular quantum computation scheme}
\label{sec:boc}

Toward the realization of oracular computations within the client-server
paradigm, while offering blindness as a security feature, we propose a
quantum computation scheme called \emph{blind oracular quantum computation}
(BOQC).  We give a sketch of the scheme here, with more mathematical
details available in~\cite{boc}.  The scheme offers a
solution in a setting with the following requirements: Alice is a client
who wants to run an oracular quantum algorithm. Oscar is another client
who is in possession of oracles and is willing to cooperate with Alice
to run her oracular algorithm. Bob is a server who owns a powerful
quantum computer on which Alice and Oscar can run the algorithm.  But
Bob is curious, and he is to be prevented (``blinded'') from acquiring
knowledge of the algorithm or its output.  For example, as previously
illustrated, in a situation when Alice wants to run a Grover algorithm,
Oscar is in the possession of database and helps Alice to discover the
marked datum in the database by implementing the Grover oracle (or its
H\o yer variant), without leaking this information to Bob or to any
other parties. 

\begin{figure}[!t]
  \centering
  \begin{minipage}{\textwidth}
    \resizebox{\columnwidth}{!}{

\begin{tikzpicture}

  \begin{scope}[font=\footnotesize,auto,every node/.style={circle,inner sep=0pt,minimum  size=7.4pt, draw=black,text width=7.5pt,align=center}]
    \node (n1) at (-.7, 2.6) {1};
    \node (n2) at (-.7, 1.2) {2};
    \node (n3) at (-.7,  .6) {3};
  \end{scope}

  \begin{scope}[>=stealth, every edge/.append
    style={decorate,decoration={snake, post length=.5mm}} ,every node/.append style={sloped,anchor=center}]

  \node (a11) at (-.2,  2.6) [abo] {A};
  \node (o11) at (1,2.6) [abo] {O};
  \draw [->, double, densely dashed](a11) -- node [midway,above]{\footnotesize$\vec{\xi}$}(o11);
  \node (k1) at (.55,2){\footnotesize$\vec r=\textsc{PRNG}(\vec\xi)$};

  \draw [dotted] (1.6, 3) -- (1.6,-.3) {};

  \node (a21) at (2,  2.6)[abo] {A};
  \node (o21) at (3.2,2.6)[abo] {O};
  \node (b21) at (2.6,1.8)[abo] {B};
  \path [->] (a21.south) edge node [above, text width=.5cm]{\tiny$\ket*{+_{\Theta_1}}$}(b21.north);

  \node (d1) at (3.8,1.8) {\dots};

  \node (a31) at (4.2,  2.6)[abo] {A};
  \node (o31) at (5.4,2.6)[abo] {O};
  \node (b31) at (4.8,1.8)[abo] {B};
  \path [->] (a31.south) edge node [above, text width=.5cm]{\tiny$\ket*{+_{\Theta_n}}$}(b31.north);

  \node (e1) at (3.8,1.2) {\footnotesize B: $\textsc{Ent}(\mathcal F_1)$};

  \draw [dotted] (5.8, 3) -- (5.8,-.3) {};

  \node (a41) at (6.2, 2.6)[abo] {A};
  \node (o41) at (7.4, 2.6)[abo] {O};
  \node (b41) at (6.8, 1.8)[abo] {B};
  \path [->] (o41.south) edge node [above, text width=1cm]{\tiny$\ket*{+_{\theta_1}}$}(b41.north);

  \node (d2) at  (8,1.8) {\dots};

  \node (a51) at (8.4, 2.6)[abo] {A};
  \node (o51) at (9.6, 2.6)[abo] {O};
  \node (b51) at (9,   1.8)[abo] {B};
  \path [->] (o51.south) edge node [above, text width=1cm]{\tiny$\ket*{+_{\theta_m}}$}(b51.north);

  \node (e1) at (8,1.2) {\footnotesize B: $\textsc{Ent}(\mathcal G_1)$};

  \draw [dotted] (10, 3) -- (10,-.3) {};

  \node (a61) at (10.4, 2.6)[abo] {A};
  \node (o61) at (11.6, 2.6)[abo] {O};
  \node (b61) at (11  , 1.8)[abo] {B};
  \path [->] (a61.south) edge node [above, text
  width=.7cm]{\tiny$\ket*{+_{\Theta_{n+1}}}$}(b61.north);

  \node (d3) at (12.2,1.8) {\dots};

  \node (a71) at (12.6, 2.6)[abo] {A};
  \node (o71) at (13.8, 2.6)[abo] {O};
  \node (b71) at (13.2, 1.8)[abo] {B};
  \path [->] (a71.south) edge node [above, text
  width=.7cm]{\tiny$\ket*{+_{\Theta_{n+l}}}$}(b71.north);

  \draw [dotted] (14.2, 3) -- (14.2,-.3) {};
  \node (e1) at (12.2,1.2) {\footnotesize B: $\textsc{Ent}(\mathcal F_2)$};
\end{scope}

\draw [-, dashed] (-.7,1.45) -- (15,1.45) {};
\draw [-, dashed] (-.7,.95) --  (15,.95) {};

\begin{scope}[>=stealth,every edge/.append style={double}, every node/.append style={sloped,anchor=center}]
  \node (a1) at (2,   .6)[abo] {A};
  \node (o1) at (3.2, .6)[abo] {O};
  \node (b1) at (2.6, 0)[abo] {B};
  \path [->](a1) edge node [above]{\tiny$\Delta_1$} (b1.north);
 \path [->] (b1.west) edge node [below]{\tiny$B_1$} (a1.south);
 \path [->] (b1.east) edge node [below]{\tiny$B_1$} (o1.south);

 \node (d1) at (3.8,0) {\dots};

 \node (a2) at (4.2,.6)[abo] {A};
 \node (o2) at (5.4,.6)[abo] {O};
 \node (b2) at (4.8,0)[abo] {B};
 \path [->] (a2) edge node [above]{\tiny$\Delta_n$} (b2.north);
 \path [->] (b2.west) edge node [below]{\tiny$B_n$} (a2.south);
 \path [->] (b2.east) edge node [below]{\tiny$B_n$} (o2.south);

 \node (a3) at (6.2,.6)[abo] {A};
 \node (o3) at (7.4,.6)[abo] {O};
 \node (b3) at (6.8,0)[abo] {B};
 \path [->] (o3) edge node [above]{\tiny$\delta_1$} (b3.north);
 \path [->] (b3.west) edge node [below]{\tiny$b_1$} (a3.south);
 \path [->] (b3.east) edge node [below]{\tiny$b_1$} (o3.south);

 \node (d2) at (8,0) {\dots};

 \node (a4) at (8.4,.6)[abo] {A};
 \node (o4) at (9.6,.6)[abo] {O};
 \node (b4) at (9, 0)[abo] {B};
 \path [->] (o4) edge node [above]{\tiny$\delta_m$} (b4.north);
 \path [->] (b4.west) edge node [below]{\tiny$b_m$} (a4.south);
 \path [->] (b4.east) edge node [below]{\tiny$b_m$} (o4.south);

 \node (a5) at (10.4,.6)[abo] {A};
 \node (o5) at (11.6, .6)[abo] {O};
 \node (b5) at (11,  0)[abo] {B};
 \path [->] (a5) edge node [above]{\tiny$\Delta_{n+1}$} (b5.north);
 \path [->] (b5.west) edge node [below]{\tiny$B_{n+1}$} (a5.south);
 \path [->] (b5.east) edge node [below]{\tiny$B_{n+1}$} (o5.south);

 \node (d3) at (12.2,0) {\dots};

 \node (a6) at (12.6,.6)[abo] {A};
 \node (o6) at (13.8,.6)[abo] {O};
 \node (b6) at (13.2, 0)[abo] {B};
 \path [->] (a6) edge node [above]{\tiny$\Delta_{n+l}$} (b6.north);
 \path [->] (b6.west) edge node [below]{\tiny$B_{n+l}$} (a6.south);
 \path [->] (b6.east) edge node [below]{\tiny$B_{n+l}$} (o6.south);

 \node (d4) at  (14.7,0) {\dots};
 \node (e1) at  (14.7,1.2) {\dots};
 \node (e1) at  (14.7,1.8) {\dots};

 \node at (1.4,1.65) {\color{gray}\footnotesize 1}; 
 \node at (5.6,1.65) {\color{gray}\footnotesize 2}; 
 \node at (5.6,1.1) {\color{gray}\footnotesize 3}; 
 \node at (5.6,-.2) {\color{gray}\footnotesize 4}; 
 \node at (9.8,1.65) {\color{gray}\footnotesize 5}; 
 \node at (9.8,1.1) {\color{gray}\footnotesize 6}; 
 \node at (9.8,-.2) {\color{gray}\footnotesize 7}; 
 \node at (14,1.65) {\color{gray}\footnotesize 8}; 
 \node at (14,1.1) {\color{gray}\footnotesize 9}; 
 \node at (14,-.2) {\color{gray}\footnotesize 10}; 
\end{scope}
\end{tikzpicture}

  } \end{minipage}

  \caption{The BOQC and the optimized BOQC protocols.  The initials A, B,
    O denote Alice, Bob, and Oscar respectively.  The double dashed
    arrow in part 1 represents an authenticated channel --- Alice sends
    Oscar a random key $\vec\xi$, which is then expanded to $\vec r$
    using a pseudorandom number generator \textsc{PRNG}($\vec\xi$).
    Double solid arrows represent classical channels, over which variables
    $\{\Delta_j,\delta_k,B_j,b_k\}$ are sent.  Wavy arrows
    represent quantum channels, where quantum states
    $\{\ket*{+_{\Theta_j}},\ket*{+_{\theta_k}}\}$ are transmitted;
    vector $\{\Theta_j\}$ is generated and known only to Alice and
    $\{\theta_j\}$ is generated and known only to Oscar.  The vertical
    dashed lines separate subgraphs. The function $\textsc{Ent}$ is an
    entangling operator performed by Bob applying CPHASE gates according
    to edges of the corresponding subgraph.  The BOQC protocol is
    performed by following these steps: \circled{1} preparation, which
    comprises key sharing and qubit transmission, \circled{2}
    entanglement, which is the applications of entangling operations by
    Bob, and \circled{3} measurement, where two-way classical
    interactions take place between clients and server; Bob measures each
    qubit in an angle instructed by Alice or Oscar, then he publicly
  announces the outcome. The optimized BOQC protocol comprises the same
steps, however following the order that is notated by grey numbers.  } 

\label{fig:boqc_protocol} 
\end{figure}
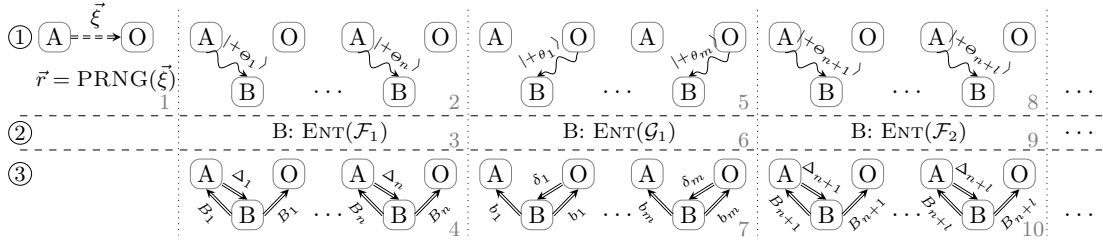%

Running a computation within the BOQC scheme comprises the following
steps.  \textbf{First}, Alice and Oscar independently plan out their
quantum computations within the 1WQC scheme. Alice will run the
non-oracle blocks and Oscar runs the oracle blocks. Let Alice's
computation be $\{(\mathcal F, I, O),\vec\Phi\}$ with input nodes $I$
and output nodes $O$, where $\mathcal F$ comprises subgraphs $\mathcal
F\equiv (\mathcal F_1,\dots,\mathcal F_p)$ together with the
corresponding angles $\vec\Phi\equiv(\vec\Phi_1,\dots,\vec\Phi_p)$. Let
Oscar's computation be $\{\mathcal G,\vec\phi\}$, where $\mathcal G$
comprises subgraphs $\mathcal G\equiv (\mathcal G_1,\dots,\mathcal G_q)$
together with the corresponding angles
$\vec\phi\equiv(\vec\phi_1,\dots,\vec\phi_q)$.  The total computation is
$\{(\mathcal T,I,O),\vec\Sigma\}$, where $\mathcal{T}\eqqcolon(\mathcal
F_1,\mathcal G_1,\mathcal F_2,\dots)$ with the corresponding angles
$\vec\Sigma\eqqcolon(\vec\Phi_1,\vec\phi_1,\vec\Phi_2,\dots)$. Alice
sends Oscar a random seed $\vec\xi$ via an authenticated channel to
share a string $\vec r$, which is generated by both of them using a
pre-agreed pseudorandom generator $\vec r=\textsc{PRNG}(\vec\xi)$, where
$\abs*{\vec{r}}\gg\abs*{\vec{\xi}}$. Then, Alice sends Bob input qubits,
followed by Alice and Oscar alternately sending Bob the remaining qubits
that correspond to non-input nodes. The non-input nodes are prepared in
a specific way: Alice prepares $\{\ket*{+_{\Theta_j}}\}$ and Oscar
prepares $\{\ket*{+_{\theta_k}}\}$, where
$\Theta_j,\theta_k\in[0,2\pi)$; $\vec\Theta$ are random angles known
only to Alice and $\vec\theta$ are random angles known only to Oscar.
{\bf Second}, Bob entangles the received qubits by applying CPHASE gates
that correspond to the edges of graph $\mathcal T$.  \textbf{Third}, Bob
measures all non-output nodes --- if there is no quantum output for
Alice, as in the Grover case, Bob measures all nodes.  These
measurements are performed one by one, where every angle is publicly
announced: Alice announces $\Delta_j\coloneqq \Theta_j+\Phi_j+\pi r_i$
if qubit $i$ corresponds to one of her nodes, Oscar announces
$\delta_k\coloneqq \theta_k+\phi_k+\pi r_i$ if qubit $i$ is one of his
nodes; Bob publicly announces his measurement outcome.  If Alice expects
quantum output, Bob sends her the output qubits in the end.  Those steps
are pictorially shown in~\autoref{fig:boqc_protocol}.

Given that $\mathcal T$ has a flow, running a computation within the BOQC
while implementing the UBQC correction scheme (see~\autoref{sec21}),
will result in the same computation as if it is run within UBQC scheme,
which is deterministic for all measurement angles and measurement
outcomes. The proof of this statement is provided in~\cite{boc}. 


If the physical qubits employed, which should have long coherence times, can be
rapidly re-initialized, the BOQC protocol can be optimized to use fewer resources;
this optimization is similar to the scheme
in~\cite{housmand2018}.  Assuming that Bob's qubits are reused
after being measured, Alice and Oscar alternately perform a complete
computation round;  pictorially this means using the order denoted with gray
numbers in~\autoref{fig:boqc_protocol}, where the last layer of
$\mathcal F_j$ is left unmeasured, becoming the input of $\mathcal
G_j$ --- thus, the last unmeasured layer of $\mathcal G_j$ is the
input of $\mathcal F_{j+1}$. 

Further optimization can be done by partitioning $\mathcal F_j$ and
$\mathcal G_j$ into smaller subgraphs, for instance, into subgraphs
comprising the qubit about to be measured and its nearest neighbors. We
demonstrate an algorithm computed within such an optimized BOQC scheme using
NV centers (\autoref{sec:nvc}). The measurement corrections of such
scheme is covered in~\cite{boc}.

We monitor the security of our protocol using the leaking function
defined in~\cite{abadi1987hiding}. We observe that the BOQC protocol is blind
while leaking only the graph structure $\mathcal{T}$.  Recall Bob possesses information
$\{\vec\Theta,\vec\theta\}\eqqcolon X$ that are attributes of the quantum states and
$\{\Delta,\delta\}\eqqcolon Y$, where $\Delta_j=\Theta_j+\Phi_j+\pi r_i$
and $\delta_k=\theta_k+\phi_k+\pi r_i$.  But the random quantities
$\{\vec\Theta,\vec\theta,\vec r\}$ are independent of actual
computation $\{\vec\Phi,\vec\phi\}$, thus $X$ is independent of $Y$.
Since $X$ is encoded within quantum states, inferring $\vec\Theta$ or
$\vec\theta$ without disturbing the quantum states is impossible. Thus,
no information is gained by Bob during protocol run --- the BOQC is
blind. 

The only catch in the security of BOQC is the establishment of the
symmetric string $\vec r$ between Alice and Oscar. We assume that Bob
cannot learn about $\vec r$ --- Bob cannot learn the random seed
$\vec\xi$ since we use an authenticated channel here.  If Bob knows the
function $\textsc{PRNG}$, guessing $\vec r$ compromises the security
with probability $2^{- \abs*{\vec{\xi}}}$.  Thus, $\vec\xi$ must be long
enough that the probability of Bob correctly guess the seed is
infinitely small.

\subsection{
Construct circuits for Grover-H\o yer algorithm}
\label{sub:constructing}

In this section, we present a strategy to obtain quantum circuits that
run the Grover-H\o yer algorithm. We will specifically explore cases
where the database is encoded within three qubits, where $N=5,6,7,$ and
$8$. The strategy essentially is seeking every circuit in $\mathcal B$
(\autoref{eq:b}) using DS94 optimization. The main challenges are the
abundance of database choices and 2-qubit gate networks to be tried out; note that
$\binom{8}{N}$ database choices are possible for each $N$. A
strategy to group those choices into a small number of equivalent sets will also be presented here.

\begin{algorithm}[h!]
  \caption{Circuit search}
  \label{alg:circ_search}
  \begin{algorithmic}[1]
  \Require $w, \tau$ 
  \State $l \gets 0$
  \State \label{line:unique}$\mathcal N\gets$ all unique topologies of
  2-qubit gates network with size $l$
  \For{$G$ in $\mathcal N$}
  \State Optimize $G$
  \If  {the optimization succeeds}

    \Return $G$, the circuit is found and \textbf{Exit}
  \EndIf 
  \EndFor
  \State $l \gets l+1$ and go to line~\ref{line:unique}
  \end{algorithmic}
\end{algorithm}

We seek quantum circuits using \Cref{alg:circ_search} --- for
$N\in\{5,6,7,8\}$, for all unique database combinations $w$, and all
marked items $\tau\in w$ --- by finding all operations in $\mathcal
B_{w,\tau}$ (see~\autoref{eq:b}), that is the required operators to run
the Grover-H\o yer algorithm for a database $w$ and a marked item
$\tau$.  Note that this search of circuits is done separately --- one
may do it for the whole Grover-H\o yer algorithm and obtain smaller
circuits ---  in order to obtain a BOQC-compatible circuit.

We will see that many database choices are equivalent by
considering the role of the Grover oracle.  
For convenience, rewrite a
database set $w=\{d_1,d_2,\dots,d_N\}\equiv d_1d_2\dots d_N$, where
$d_j\in \{0,1,2,3,4,5,6,7\}$.
Given three bits $\ket{ijk}$ to encode $w$, where
$i,j,k\in\{0,1\}$, and a set of oracle operators where each of them
``marks'' one element by phase $e^{i\varphi}$.  Two sets of database
$w_1,w_2$, where $\abs*{w_1}=\abs*{w_2}$, are equivalent if a set of
oracles that can mark for all $\tau_1\in w_1$ can also mark for all
$\tau_2\in w_2$ up to some global phases. 

By this means, while
considering their bit representations, $w_1$ is equivalent to $w_2$ if
they are identical up to permutation and bit complementation. For
instance, consider two equivalent databases with their bit
representations (in little-endian format):
$01234=\{000,001,010,011,100\}, 10543=\{001,000,101,100,011\}$. One
can be obtained from another by complementing the third bit and permuting
the first and the second bits. 

In the gate model, it
means their oracles are equivalent up to some operations:
  \begin{equation}
      \small\Qcircuit @C=.5em @R=.5em {
 &\multigate{2}{O_{01234}}&\qw & & & & &\qswap&\multigate{2}{O_{10543}}    &\qswap&\qw \\
 &\ghost{O_{01234}} &\qw&&\equiv&&&\qswap\qwx&\ghost{O_{10543}} &\qswap\qwx&\qw&,\\
 &\ghost{O_{01234}} &\qw&& &&&\gate{X}&\ghost{O_{10543}}       &\gate{X}&\qw \\
}
\end{equation}
where $O_{01234}$ and $O_{10543}$ represents the oracle operators for
databases $01234$ and $10543$ respectively. With these equivalences, all databases are covered by the following set: 
\begin{equation}
\{01234, 01247, 01256,
012345, 012347, 012567, 0123456, 01234567\}\eqqcolon \mathcal D.
  \label{eq:sb}
\end{equation}
Note that this strategy works for an arbitrary number of bits, not
only for three.

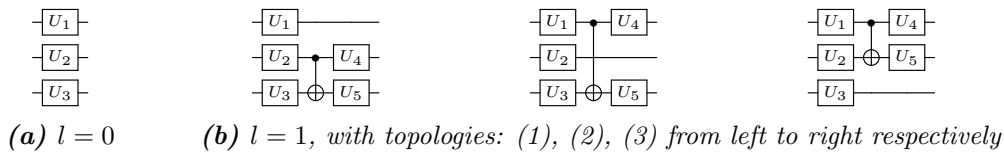
\begin{figure}[t!]
  \centering
  \begin{subfigure}[b]{0.2\textwidth}
    \begin{minipage}[b]{\textwidth}
    \[ \tiny\Qcircuit @C=.5em @R=.5em {
      & \gate{U_1} & \qw \\
      & \gate{U_2} & \qw \\
      & \gate{U_3} & \qw  }\]
    \end{minipage}
    \caption{$l=0$}
    \label{fig:network_example_l0}
  \end{subfigure}
  \begin{subfigure}[b]{0.75\textwidth}
    \begin{minipage}[b]{0.32\textwidth}
      \[\tiny \Qcircuit @C=.5em @R=.5em {
      & \gate{U_1} & \qw     &\qw & \qw \\
      & \gate{U_2} & \ctrl{1}&\gate{U_4} & \qw \\
      & \gate{U_3} & \targ   &\gate{U_5}& \qw 
    }\]
    \end{minipage}
    \begin{minipage}[b]{0.32\textwidth}
      \[\tiny \Qcircuit @C=.5em @R=.5em {
      & \gate{U_1}&\ctrl{2}&\gate{U_4}&\qw \\
      & \gate{U_2}&\qw     & \qw      &\qw \\
      & \gate{U_3}&\targ   &\gate{U_5}&\qw 
    }\]
    \end{minipage}
    \begin{minipage}[b]{0.32\textwidth}
      \[\tiny\Qcircuit @C=.5em @R=.5em {
      & \gate{U_1} & \ctrl{1}&\gate{U_4} & \qw \\
      & \gate{U_2} & \targ   &\gate{U_5} & \qw \\
      & \gate{U_3} & \qw     & \qw& \qw 
    } \]
    \end{minipage}
    \caption{$l=1$, with topologies: (1), (2), (3)
    from left to right respectively}
    \label{fig:network_example_l1}
  \end{subfigure}

  \caption{The distinct 2-qubit gate networks for $l=0$ and $l=1$.
   Operators $U_j\equiv U_j(\alpha_j,\beta_j,\gamma_j)$ are
 1-qubit gates as in~\autoref{eq:oneunitary}.}
  \label{fig:network_example}
\end{figure}

As one may freely define a set of quantum gates that compose gate
networks (for instance DS94 considered the set of all U(4) matrices), we follow 1WQC and compose the gate networks into the operations $\{$CNOT,
$U\}$, where $U$ is a unitary matrix in SU(2) having the form
\begin{equation} 
  \label{eq:oneunitary}
  U(\alpha,\beta,\gamma)= \begin{pmatrix}
e^{i{\beta}}\cos(\alpha) & e^{i{\gamma}}\sin(\alpha) \\
-e^{-i{\gamma}}\sin(\alpha) & e^{-i{\beta}}\cos(\alpha) &
\end{pmatrix},
\end{equation} 
where $\alpha,\beta,\gamma\in [0,2\pi)$ are free parameters; these will be the optimization parameters below. We define $l$ to be the number of CNOTs in our three-qubit network.   For $l=0$, the network is simply three 1-qubit
gates; for every additional CNOT gate, four 1-qubit gates
are added, two before and two after. Thus, $6l+9$ free parameters
will be available for the optimization for a network with size $l$. All networks for
$l=0$ and $l=1$ are shown in~\autoref{fig:network_example}. 

Again, not all networks are distinct; we obtain a minimal set of representative networks by
enumerating all possible network topologies, followed by two
eliminations: we eliminate ones that have more than three consecutive CNOT
gates, and we eliminate the ones that are topologically
equivalent~\cite{divincenzo1994results} (also discussed
in~\Cref{sub:exhaustive}).  The first elimination is based on the
fact that an arbitrary SU(8) can be constructed using three CNOT gates and eight 1-qubit
gates~\cite{vidal2004universal}.  Thus, for example, topology $(133331)$
is eliminated since $(133331)=(13331)$.

We use DS94 optimization within~\Cref{alg:circ_search} to find the gate
networks. A BFGS solver of the Python SciPy library~\cite{scipy} is
employed in our program. To speed up optimizations, we define
more relaxed objective functions than DS94:
\begin{equation} f_b=\sum_{i,j \in w,M_{ij}\neq 0}\abs{M_{ij}-S_{ij}}^2
  \quad\quad  f_p=\sum_{i\in w, M_{i0}\neq
0}\abs{M_{i0}-S_{i0}}^2,\label{eq:objective_fpb} \end{equation} where
$M$ is the desired unitary matrix and $S$ is the resulting matrix from
the tested network ($G$); $f_p$ is used if $M=A$, that is the
preparation block in $\mathcal B_{w,\tau}$, and $f_b$ is used for other
blocks in $\mathcal B_{w,\tau}\setminus \{A\}$.  While the $M$s are assumed to
be unitary matrices here, it is sufficient to consider only the non-zero
elements within the subspace that is induced by $w$. Note that $f_p$ is
appropriate for $M=A$ because we start from the all-zero state $\ket 0^{\otimes n}$.

Success in optimization is defined as $f_p \leq \varepsilon$ or $f_b
\leq \varepsilon$, where $\varepsilon$ is a chosen
numerical precision. We define $\varepsilon$ such that the success
probability is approximately one: given $\delta$, there exists an
$\varepsilon$ such that $p_s \ge 1-\delta$, where $p_s$ is the success
probability of running~\Cref{alg:exactgrover} while replacing block $M$
with the tested network $G$.  For the non-oracle cases, $M\neq O$, we
take the worst $p_s$ among all obtained $p_s$ from different marked
items.

\begin{table}[!t]
  \centering
  \begin{subtable}{.46\linewidth}
    \caption{\footnotesize $N$=5, $\varphi+u$=0.1707, $\psi$=0.4510}
    \label{tab:countcx5}
\resizebox{\columnwidth}{!}{
  \begin{tabular}{|c|c|ccccc|c|ccccc|c|}\hline
$w$& $\mathcal{A}$ &\multicolumn{5}{c|}{$O(\pi)$}&
$D(\pi)$ &\multicolumn{5}{c|}{$O(\varphi+u)$}& $D(\psi)$  \\
\hline
01234  & &
0& 1& 2& 3& 4& & 0&1&2&3&4&\\[-3pt]
\cline{3-7}\cline{9-13}
\small{CNOT} &2
  &1&1&1&1&0& 7
  &2&2&2&2&0& 8   \\
\hline
01247  & &
0&1&2&4&7& & 0&1&2&4&7&\\[-3pt]
\cline{3-7}\cline{9-13}
\small{CNOT} &2
  &0&1&1&1&1& 7
  &0&2&2&2&2& 8   \\
\hline
01256  & &
0&1&2&5&6& & 0&1&2&5&6&\\[-3pt]
\cline{3-7}\cline{9-13}
\small{CNOT} &3
  &0&1&1&1&1& 8
  &0&2&2&2&2& 9  \\ 
\hline
\end{tabular}
}
\end{subtable}%
\hfill
\begin{subtable}{.52\linewidth}
  \caption{\footnotesize $N$=6, $\varphi+u$=1.861, $\psi$=0.841}
  \label{tab:countcx6}
\resizebox{\columnwidth}{!}{%
\begin{tabular}{|c|c|cccccc|c|cccccc|c|}\hline
$w$& $\mathcal{A}$ &\multicolumn{6}{c|}{$O(\pi)$}&
$D(\pi)$ &\multicolumn{6}{c|}{$O(\varphi+u)$}& $D(\psi)$  \\
\hline
\bfseries 012345  & &
\bfseries 0&\bfseries1&\bfseries2&\bfseries3&\bfseries4&\bfseries5& & \bfseries0&\bfseries1&\bfseries2&\bfseries3&\bfseries4&\bfseries5&\\[-3pt]
\cline{3-8}\cline{10-15}
\bfseries\small{CNOT} & \bfseries1
  &\bfseries2&\bfseries2&\bfseries1&\bfseries1&\bfseries1&\bfseries1&\bfseries 4
  &\bfseries4&\bfseries4&\bfseries2&\bfseries2&\bfseries2&\bfseries2&\bfseries 6   \\
\hline
012347  & &
0&1&2&3&4&7& & 0&1&2&3&4&7&\\[-3pt]
\cline{3-8}\cline{10-15}
\small{CNOT}    &2
  &2&1&1&2&1&1& 6
  &4&2&2&4&2&2& 7   \\
\hline
012567  & &
0&1&2&5&6&7& & 0&1&2&5&6&7&\\[-3pt]
\cline{3-8}\cline{10-15}
\small{CNOT}    &3
  &1&1&1&2&1&1& 8
  &2&2&2&2&2&2& 8   \\
\hline
\end{tabular}
}
\end{subtable} 

\begin{subtable}{.65\linewidth}
\caption{\footnotesize $N$=7, $\varphi+u$=2.0277, $\psi$=1.2056}
\label{tab:countcx7}
\resizebox{\columnwidth}{!}{%
\begin{tabular}{|c|c|ccccccc|c|ccccccc|c|}\hline
$w$& $\mathcal{A}$ &\multicolumn{7}{c|}{$O(\pi)$}&
$D(\pi)$ &\multicolumn{7}{c|}{$O(\varphi+u)$}& $D(\psi)$  \\
\hline
0123456  & &
0&1&2&3&4&5&6& & 0&1&2&3&4&5&6&\\[-3pt]
\cline{3-9}\cline{11-17}
\small{CNOT}  & 3
  &3&2&2&1&2&1&1& 8
  &4&4&6&2&4&2&2& 9   \\
\hline
\end{tabular}
}
\end{subtable} 

\begin{subtable}{.7\linewidth}
\caption{\footnotesize $N$=8, $\varphi+u$=2.2143, $\psi$=1.5708}
\label{tab:countcx8}
\resizebox{\columnwidth}{!}{%
\begin{tabular}{|c|c|cccccccc|c|cccccccc|c|}\hline
$w$& $\mathcal{A}$ &\multicolumn{8}{c|}{$O(\pi)$}&
$D(\pi)$ &\multicolumn{8}{c|}{$O(\varphi+u)$}& $D(\psi)$  \\
\hline
01234567  & &
0&1&2&3&4&5&6&7& & 0&1&2&3&4&5&6&7&\\[-3pt]
\cline{3-10}\cline{12-19}
\small{CNOT}  &   0
  &6&6&6&6&6&6&6&6& 6
  &6&6&6&6&6&6&6&6& 6   \\
\hline
\end{tabular}
}
\end{subtable} 
\caption{The number of CNOT gates for all distinct combinations of
  $(w,\tau,M)$ for all $w\in \mathcal D$, $\tau\in w$,
  $M\in \mathcal B_w$, and
$N\in\{5,6,7,8\}$. The angles $\varphi,\psi$, and $u$ refer to the H\o yer
exact-Grover technique, see~\Cref{alg:exactgrover}.}

\label{tab:CNOT_counting}
\end{table}

We obtain~\Cref{tab:CNOT_counting}, which shows the size of the network
for every operator in $\mathcal B_{w,\tau}$, for all unique database
sets $w\in \mathcal D$, and for all valid marked items, where 
$\delta = 10^{-4}$.  We obtain $\varepsilon\leq 4.8\times 10^{-11}$ for
preparation blocks and $\varepsilon_b \leq 2.5 \times 10^{-8}$ for other
blocks.  The complete tables that show values of success probabilities
$p_s$ are shown in the Appendix, (\Cref{tab:prob}). While this does not
complete our analysis of the three-qubit Grover algorithms, these
preliminary calculations indicate that the most efficient network will
be achieved for $N=6$ and $w=012345$ (and not the smaller $N=5$).

At this point, we complete the classical processing stage
of~\Cref{alg:exactgrover}. Since the blocks are prepared independently,
this result can be adapted to develop the full BOQC scheme.  However,
for $N<8$, the straightforward implementation of the oracles would
require different network sizes for different marked items $\tau$.  This
would allow Bob to learn about Alice's request to Oscar.  In the next
section, we complete our exact quantum search algorithm, taking care
that networks of identical structure are created for each value of
$\tau$, assuring the blindness of the protocol.

\subsection{The exact quantum search algorithm with blind oracles}
\label{sub:exactblind}

\begin{algorithm}[t!]
\caption{Blind exact quantum search}
\label{alg:exactgrover_blind}
\begin{algorithmic}[1]
  \Require $n,w,\tau, M_{POVM}$  

  \Statex{\bfseries (1.a) Classical processing done by Alice }

  \State{ Prepare $\{\mathcal A, \mathcal D(\pi),\mathcal D(\psi)\}$
    (\autoref{eq:b}) using~\Cref{alg:circ_search}, with objective
    functions $f_a$ or $f_b$ (\autoref{eq:objective_fpb}).}

  \Statex{\bfseries (1.b) Classical processing done by Oscar }

  \State {Search for the blind oracles $\tilde{\mathcal O}$
    using~\Cref{alg:circ_search} with the objective function
    \[\textsc{ObjPOVM}(G_j,M_{POVM},\vec{\varphi},n,m,w,A,D(\pi),D(\psi)),\] 
defined in~\cref{subr:obj_povm}, where $G_j$ is the tested network,
$\vec \varphi\equiv (\vec \varphi_1,\dots,\vec\varphi_N)$ is the
optimization parameters --- with random initialization --- for all
marked items $\tau\in w$, $M_{POVM}$ is the defined POVM, and the rest
($n,m,w,A,D(\pi),D(\psi)$) are defined as those in~\Cref{alg:exactgrover}.
If the optimization succeeds, set $\tilde{\mathcal O}\leftarrow G_j$ and
$\vec \varphi$ is optimized.  All oracles now have the same networks,
but different parameters, which are set according to the marked item. 
 } 
\State{Set $\mathcal O(\pi)\gets\tilde{\mathcal O}$ and  $\mathcal
  O(\varphi+u)\gets\tilde{\mathcal O}$;
    the two oracle calls labelled $\mathcal O(\pi)$ and
    $\mathcal O(\varphi +
  u)$ in previous algorithms will be accomplished by the same oracle operation $\tilde{\mathcal
O}$.}
\Statex {\bfseries (2) Quantum processing}
\State{Perform the quantum processing of~\Cref{alg:exactgrover} within
  the respective scheme. If the scheme is BOQC, run it according to the scheme of~\autoref{fig:boqc_protocol}.}
\State{\bfseries Exit}
\end{algorithmic}
\end{algorithm}
\begin{subroutine}[hb!]
\caption{Objective function based on POVM}
\label{subr:obj_povm}
\begin{algorithmic}[1]
  \Function{ObjPOVM}{$G_j,M_{POVM},\vec{\varphi},n,m,w,A,D(\pi),D(\psi)$}
    \State $ov\gets 0$, the objective value 
    \For {$\tau$ in $w$\label{line:loop}}
    \State $S\gets G_j(\vec{\varphi}_{\tau}$)
    \State $\ket*\Psi\gets A\ket0^{\otimes n}$ 
      \For{$i=1$ to $m$}
      \State $\ket*\Psi\gets D(\pi)S\ket*\Psi$ 
      \EndFor
      \State $\ket*\Psi\gets D(\psi)S\ket*\Psi$
      \State\label{line:c1}$ov\gets ov + 1-\abs{\bra{\tau}M_{POVM}(\tau)\ket*\Psi}$ 
      \State\label{line:c2}$ov\gets ov + \abs{f_{obd}}$ (\autoref{eq:objective_foffd}) 
    \EndFor
    \State \Return $ov$
\EndFunction
  \end{algorithmic}
\end{subroutine}

Here we introduce an algorithm called \emph{blind exact quantum search
algorithm} (BEQS), given in~\Cref{alg:exactgrover_blind}, which is an
improvement of the Grover-H\o yer algorithm: it is compatible with the
BOQC scheme, and it involves a general scheme for storing information in
the database. Achieving the first means obtaining identical oracles for
all marked items, whose measurement angles are adjusted accordingly.
The latter means permitting several marked items $\tau$ to stand for a
single database entry; we do this by making the final measurement of the
Grover algorithm an incomplete or POVM measurement. For instance,
consider 2-bit Grover algorithm with database $w=0123$, $N=4$, and
database entries \{A,B\}  ($\tilde N=2$). Note that here we distinguish
between the database size ($N$) and the number of database entries
($\tilde N$).  Alice and Oscar agree ahead of time that either outcome 0
or 1 correspond to entry A, and outcome 2 or 3 correspond to entry B;
this is attainable by defining measurement operators
$\{\ketbra0+\ketbra1,\ketbra2+\ketbra3\}$.  We will refer to such a
scheme as ``POVM measurement strategy''.  Physically, it is possible to
still do the full projective measurement, then classically associate the
measurement outcomes with database entries.

While it is hard analytically to obtain an identical form --- in our
case using gate networks --- of oracles for all marked items, the BEQS
provides a numerical method to obtain all those oracles using a single
numerical procedure. The key lies in the objective function
\textsc{ObjPOVM} in~\Cref{subr:obj_povm}, which includes two
constraints: \textbf{(C1)} the success probability must be one, and
\textbf{(C2)} the resulting operator must preserve the state space.
Recall that database choice $w$ induces the state space.

The first constraint, \textbf{C1}, implemented at line~\ref{line:c1}
of~\Cref{subr:obj_povm}, imposes a successful computation within the
defined POVM measurement for all permitted marked items --- notice that
the loop goes for all $\tau\in w$.  Constraint \textbf{C2}, implemented
at line~\ref{line:c2} of~\Cref{subr:obj_povm}, assures a block diagonal
matrix, which is critical when there is a free subspace in the full
$2^n$-dimensional Hilbert
space for an $n$-qubit system. This constraint is imposed by requiring that the sum of the
absolute values of the elements outside diagonal block to be zero:
\begin{equation}
  f_{obd} = \sum_{i\in x, j\in x\setminus w,  i\neq j
  }\abs{S_{ij}}+ \sum_{i\in x\setminus w,j\in w} \abs{S_{ij}},
  \label{eq:objective_foffd}
\end{equation}
where $w$ is the database index, $x$ are all possible
indices that can be accommodated, and $S$ is the matrix from
evaluating a network $G_j(\vec \varphi_\tau)$. 
All constraints are quantified within the objective value $ov$. It is worth
mentioning that the obtained operator $\tilde O$ forms a block diagonal
matrix that does necessarily resemble neither the Grover nor the H\o yer oracles.

\begin{figure}[t!]
  \centering
\begin{minipage}[hbt!]{\textwidth}
\resizebox{\columnwidth}{!}{
  \Qcircuit @C=0.5em @R=1.200000em{
&&\gate{U_{1}}&\ctrl{1}&\gate{U_{4}}&\qw&\qw&\gate{U_{8}}&\ctrl{1}&\gate{U_{11}}&\ctrl{1}&\gate{U_{13}}&\gate{U_{15}}&\ctrl{1}&\gate{U_{18}}&\ctrl{2}&\gate{U_{20}}&\qw&\qw&\ctrl{1}&\gate{U_{24}}&\qw&\qw&\gate{U_{8}}&\ctrl{1}&\gate{U_{11}}&\ctrl{1}&\gate{U_{13}}&\gate{U_{26}}&\qw&\qw&\ctrl{1}&\gate{U_{31}}&\ctrl{2}&\gate{U_{33}}&\qw&\qw&\ctrl{1}&\gate{U_{37}}&\ctrl{2}&\gate{U_{39}}&\qw\\ 
&&\gate{U_{2}}&\targ\qw&\gate{U_{5}}&\gate{U_{6}}&\ctrl{1}&\gate{U_{9}}&\targ\qw&\gate{U_{12}}&\targ\qw&\gate{U_{14}}&\gate{U_{16}}&\targ\qw&\gate{U_{19}}&\qw&\qw&\ctrl{1}&\gate{U_{22}}&\targ\qw&\gate{U_{25}}&\gate{U_{6}}&\ctrl{1}&\gate{U_{9}}&\targ\qw&\gate{U_{12}}&\targ\qw&\gate{U_{14}}&\gate{U_{27}}&\ctrl{1}&\gate{U_{29}}&\targ\qw&\gate{U_{32}}&\qw&\qw&\ctrl{1}&\gate{U_{35}}&\targ\qw&\gate{U_{38}}&\qw&\qw&\qw\\ 
&&\gate{U_{3}}&\qw&\qw&\gate{U_{7}}&\targ\qw&\gate{U_{10}}&\qw&\qw&\qw&\qw&\gate{U_{17}}&\qw&\qw&\targ\qw&\gate{U_{21}}&\targ\qw&\gate{U_{23}}&\qw&\qw&\gate{U_{7}}&\targ\qw&\gate{U_{10}}&\qw&\qw&\qw&\qw&\gate{U_{28}}&\targ\qw&\gate{U_{30}}&\qw&\qw&\targ\qw&\gate{U_{34}}&\targ\qw&\gate{U_{36}}&\qw&\qw&\targ\qw&\gate{U_{40}}&\qw
\gategroup{1}{3}{3}{5}{.3em}  {--}
\gategroup{1}{6}{3}{12}{.3em} {--}
\gategroup{1}{13}{3}{21}{.3em}{--}
\gategroup{1}{22}{3}{28}{.3em}{--}
\gategroup{1}{29}{3}{41}{.3em}{--}\\
&&&\mathcal A&
&&&&\tilde O&&&
&&&&&D(\pi)&&&&
&&&&\tilde O &&&
&&&&&&& D(\psi)&&&&&&}}
\end{minipage}\vspace{1em}
\begin{minipage}{0.192\linewidth}
\resizebox{\columnwidth}{!}{
\begin{tabular}{|c|c|c|c|}
\hline
\textbf{Gate}&$\alpha$&$\beta$&$\gamma$\\
\hline
$U_{1}$ &2.1863&3.4700&-2.8132\\
$U_{2}$ &-3.7673&-2.8895&2.1618\\
$U_{3}$ &-0.7854&-1.5708&1.5708\\
$U_{4}$ &1.5708&1.8160&-1.2424\\
$U_{5}$ &-0.6248&-3.3919&-0.9832\\
$U_{15}$ &0.3929&1.9270&1.9237\\
$U_{16}$ &0.7011&2.8271&1.0836\\
\hline
\end{tabular}
}
\end{minipage}
\begin{minipage}{0.192\linewidth}
\resizebox{\columnwidth}{!}{
\begin{tabular}{|c|c|c|c|}
\hline
$U_{17}$ &-1.5708&-3.5218&1.5708\\
$U_{18}$ &-1.2628&1.3905&-1.0366\\
$U_{19}$ &0.8541&-3.5655&2.1142\\
$U_{20}$ &2.8341&0.3039&-2.4836\\
$U_{21}$ &1.5708&-1.9918&1.5708\\
$U_{22}$ &-0.7643&-0.6062&-1.9207\\
\hline
\end{tabular}
}
\end{minipage}
\begin{minipage}{0.192\linewidth}
\resizebox{\columnwidth}{!}{
\begin{tabular}{|c|c|c|c|}
\hline
$U_{23}$ &0.0000&-3.1416&-2.2301\\
$U_{24}$ &1.9641&2.6566&0.4818\\
$U_{25}$ &-2.3822&1.9380&-0.4206\\
$U_{26}$ &-1.9683&-0.4832&2.8293\\
$U_{27}$ &-2.1083&1.0410&-1.4486\\
$U_{28}$ &-2.9313&-1.5708&0.0000\\
\hline
\end{tabular}
}
\end{minipage}
\begin{minipage}{0.192\linewidth}
\resizebox{\columnwidth}{!}{
\begin{tabular}{|c|c|c|c|}
\hline
$U_{29}$ &-0.8967&0.4235&2.8814\\
$U_{30}$ &2.5762&0.2296&-1.4368\\
$U_{31}$ &-3.9759&-1.2708&-1.5393\\
$U_{32}$ &1.5708&-2.9348&-3.1416\\
$U_{33}$ &1.9640&-1.7740&-2.7122\\
$U_{34}$ &0.9027&-1.7681&-1.6960\\
\hline
\end{tabular}
}
\end{minipage}
\begin{minipage}{0.192\linewidth}
\resizebox{\columnwidth}{!}{
\begin{tabular}{|c|c|c|c|}
\hline
$U_{35}$ &2.2493&2.3631&0.7364\\
$U_{36}$ &3.1416&3.1416&-1.2192\\
$U_{37}$ &-0.6940&-1.7315&-1.2317\\
$U_{38}$ &1.0992&1.0522&2.5764\\
$U_{39}$ &-2.0314&-0.6042&-0.1570\\
$U_{40}$ &1.4576&1.4573&-1.8297\\
\hline
\end{tabular}}
\end{minipage}

\begin{minipage}{0.16\linewidth}
\resizebox{\columnwidth}{!}{
\begin{tabular}{|c|c|c|c|}
\hline
\multicolumn{4}{|c|}{$\tau=$ 0}\\
\hline
$U_{6}$ &-0.0000&-1.0565&2.4690\\
$U_{7}$ &-0.0000&-1.5708&2.3946\\
$U_{8}$ &0.8805&-1.2429&1.9146\\
$U_{9}$ &-3.3566&0.4472&2.0889\\
$U_{10}$ &-1.5708&-1.6829&1.5708\\
$U_{11}$ &-1.0405&0.2804&1.5793\\
$U_{12}$ &-1.3547&-2.8181&-2.0741\\
$U_{13}$ &2.5244&-2.1099&-1.6613\\
$U_{14}$ &-2.4523&-2.1062&2.3886\\
\hline
\end{tabular}}
\end{minipage}
\begin{minipage}{0.16\linewidth}
\resizebox{\columnwidth}{!}{
\begin{tabular}{|c|c|c|c|}
\hline
\multicolumn{4}{|c|}{$\tau=$ 1}\\
\hline
$U_{6}$ &0.0000&2.2951&1.0317\\
$U_{7}$ &-1.5708&-1.7305&3.1416\\
$U_{8}$ &-3.1416&-1.8576&-3.1566\\
$U_{9}$ &1.5234&1.1716&0.8384\\
$U_{10}$ &-3.1416&-1.5708&-2.5474\\
$U_{11}$ &-0.0000&-2.2763&2.5518\\
$U_{12}$ &-3.5207&2.4785&-2.1300\\
$U_{13}$ &0.0000&-1.1946&2.9811\\
$U_{14}$ &-1.1506&-2.0590&0.7998\\
\hline
\end{tabular}}
\end{minipage}
\begin{minipage}{0.16\linewidth}
\resizebox{\columnwidth}{!}{
\begin{tabular}{|c|c|c|c|}
\hline
\multicolumn{4}{|c|}{$\tau=$ 2}\\
\hline
$U_{6}$ &1.5708&-0.7819&2.2216\\
$U_{7}$ &0.1287&-3.0091&-0.5770\\
$U_{8}$ &-2.1183&-2.7515&-0.6395\\
$U_{9}$ &2.7877&-0.5423&-0.5822\\
$U_{10}$ &-0.1287&2.4033&-1.1829\\
$U_{11}$ &1.9797&-0.6641&1.5769\\
$U_{12}$ &-1.3631&2.2995&2.3634\\
$U_{13}$ &-0.5475&-1.8379&-3.4156\\
$U_{14}$ &-4.5199&0.0188&-2.3964\\
\hline
\end{tabular}}
\end{minipage}
\begin{minipage}{0.16\linewidth}
\resizebox{\columnwidth}{!}{
\begin{tabular}{|c|c|c|c|}
\hline
\multicolumn{4}{|c|}{$\tau=$ 3}\\
\hline
$U_{6}$ &1.5708&0.7393&1.9242\\
$U_{7}$ &-0.8921&3.3728&2.8631\\
$U_{8}$ &2.4257&0.0141&1.7221\\
$U_{9}$ &0.3262&-2.8357&-1.6128\\
$U_{10}$ &-0.8921&1.5255&-1.6635\\
$U_{11}$ &-0.6357&2.6774&1.5045\\
$U_{12}$ &-2.1376&-2.0598&-2.1230\\
$U_{13}$ &0.6915&1.0830&2.5152\\
$U_{14}$ &-0.0642&0.9305&-2.9311\\
\hline
\end{tabular}}
\end{minipage}
\begin{minipage}{0.16\linewidth}
\resizebox{\columnwidth}{!}{
\begin{tabular}{|c|c|c|c|}
\hline
\multicolumn{4}{|c|}{$\tau=$ 4}\\
\hline
$U_{6}$ &0.0000&1.8189&-1.8816\\
$U_{7}$ &1.7606&-0.4992&-0.4738\\
$U_{8}$ &2.3289&0.1070&-3.2364\\
$U_{9}$ &2.9211&1.0980&-2.6299\\
$U_{10}$ &-0.1898&-0.6716&3.4430\\
$U_{11}$ &1.9619&1.6645&2.0202\\
$U_{12}$ &3.6540&3.1061&0.6672\\
$U_{13}$ &-0.7716&-1.0469&-0.3914\\
$U_{14}$ &-1.5381&-1.9473&0.4896\\
\hline
\end{tabular}}
\end{minipage}
\begin{minipage}{0.16\linewidth}
\resizebox{\columnwidth}{!}{
\begin{tabular}{|c|c|c|c|}
\hline
\multicolumn{4}{|c|}{$\tau=$ 5}\\
\hline
$U_{6}$ &0.0000&2.7690&1.7284\\
$U_{7}$ &-1.0390&-2.6136&-0.6619\\
$U_{8}$ &-2.4657&-0.1839&-2.1417\\
$U_{9}$ &0.3776&-1.1476&-0.3867\\
$U_{10}$ &3.6734&-3.3307&-0.0552\\
$U_{11}$ &-3.5873&-2.7614&0.4414\\
$U_{12}$ &1.2797&-0.5862&2.5852\\
$U_{13}$ &-0.7099&0.2160&0.9320\\
$U_{14}$ &1.6530&0.2316&0.0389\\
\hline
\end{tabular}}
\end{minipage}
\caption{A 3-qubit BEQS (see~\Cref{alg:exactgrover_blind}) within the
  gate model for $w=012345$ and
$M_{POVM}=\{\ketbra0+\ketbra1,\ketbra2,\ketbra3,\ketbra4,\ketbra5\}$.
Outcomes 0 and 1 refer to the same data, thus $N=6$ and $\tilde N=5$. The circuit is
composed with \{CNOT, $U$\}, where $U$ is a SU(2) matrix and has the form
of~\autoref{eq:u}.  The parameters for oracles --- shown in the tables
below ---  are different for each marked item $\tau$.  The circuit has
success probabilities $p_s\geq 1-10^{-4}$ for all marked items $\tau\in w$.}
  \label{fig:circuit6}
\end{figure}

\begin{figure}[!t]
  \centering
  \begin{minipage}{\textwidth}
    \resizebox{\columnwidth}{!}{
      \begin{tikzpicture}
[font=\tiny,scale=.5,auto,every node/.style={circle,inner sep=0pt,minimum size=7.4pt,draw=black,text width=7.5pt,align=center}]
\node[] (n1) at (0,2){2};
\node[] (n2) at (0,1){1};
\node[] (n3) at (0,0){5};
\node[] (n4) at (1,2){3};
\node[] (n5) at (1,1){4};
\node[] (n6) at (2,2){6};
\node[] (n7) at (2,1){8};
\node[] (n8) at (3,2){9};
\node[] (n9) at (3,1){10};
\node[] (n10) at (4,2){11};
\node[red] (n11) at (5,2){14};
\node[red] (n12) at (5,0){7};
\node[red] (n13) at (6,2){16};
\node[red] (n14) at (6,0){12};
\node[red] (n15) at (5,1){13};
\node[red] (n16) at (7,0){23};
\node[red] (n17) at (6,1){15};
\node[red] (n18) at (8,0){27};
\node[red] (n19) at (7,2){18};
\node[red] (n20) at (7,1){17};
\node[red] (n21) at (8,2){19};
\node[red] (n22) at (8,1){20};
\node[red] (n23) at (9,2){22};
\node[red] (n24) at (9,1){21};
\node[red] (n25) at (9,0){31};
\node[red] (n26) at (10,2){25};
\node[red] (n27) at (10,1){24};
\node[red] (n28) at (11,2){26};
\node[red] (n29) at (11,1){28};
\node[] (n30) at (12,2){29};
\node[] (n31) at (12,1){30};
\node[] (n32) at (13,2){32};
\node[] (n33) at (13,1){34};
\node[] (n34) at (13,0){33};
\node[] (n35) at (14,2){36};
\node[] (n36) at (14,1){35};
\node[] (n37) at (14,0){37};
\node[] (n38) at (15,2){39};
\node[] (n39) at (15,1){38};
\node[] (n40) at (15,0){41};
\node[] (n41) at (16,2){40};
\node[] (n42) at (16,1){42};
\node[] (n43) at (17,2){45};
\node[] (n44) at (17,1){43};
\node[] (n45) at (18,2){47};
\node[] (n46) at (18,1){46};
\node[red] (n47) at (19,2){50};
\node[red] (n48) at (19,0){44};
\node[red] (n49) at (20,2){51};
\node[red] (n50) at (20,0){48};
\node[red] (n51) at (19,1){49};
\node[red] (n52) at (21,0){59};
\node[red] (n53) at (20,1){52};
\node[red] (n54) at (22,0){62};
\node[red] (n55) at (21,2){53};
\node[red] (n56) at (21,1){54};
\node[red] (n57) at (22,2){56};
\node[red] (n58) at (22,1){55};
\node[red] (n59) at (23,2){61};
\node[red] (n60) at (23,1){57};
\node[red] (n61) at (23,0){65};
\node[red] (n62) at (24,2){64};
\node[red] (n63) at (24,1){58};
\node[red] (n64) at (25,2){67};
\node[red] (n65) at (25,1){60};
\node[] (n66) at (26,2){70};
\node[] (n67) at (26,1){63};
\node[] (n68) at (27,2){71};
\node[] (n69) at (27,1){66};
\node[] (n70) at (27,0){68};
\node[] (n71) at (28,1){69};
\node[] (n72) at (28,0){74};
\node[] (n73) at (29,1){72};
\node[] (n74) at (29,0){76};
\node[] (n75) at (28,2){73};
\node[] (n76) at (30,1){77};
\node[] (n77) at (29,2){75};
\node[] (n78) at (31,1){79};
\node[] (n79) at (30,2){82};
\node[] (n80) at (30,0){78};
\node[] (n81) at (31,2){83};
\node[] (n82) at (31,0){80};
\node[] (n83) at (32,1){81};
\node[] (n84) at (32,0){85};
\node[] (n85) at (33,1){84};
\node[] (n86) at (33,0){88};
\node[] (n87) at (32,2){86};
\node[] (n88) at (34,1){90};
\node[] (n89) at (33,2){87};
\node[] (n90) at (35,1){92};
\node[] (n91) at (36,1){97};
\node[] (n92) at (34,2){89};
\node[] (n93) at (34,0){91};
\node[] (n94) at (35,2){93};
\node[] (n95) at (35,0){94};
\node[] (n96) at (36,2){95};
\node[] (n97) at (36,0){96};
\draw[](n1) -- (n4);
\draw[](n2) -- (n5);
\draw[](n3) -- (n12);
\draw[](n4) -- (n6);
\draw[](n5) -- (n7);
\draw[](n6) -- (n8);
\draw[](n6) -- (n7);
\draw[](n7) -- (n9);
\draw[](n8) -- (n10);
\draw[](n9) -- (n15);
\draw[](n10) -- (n11);
\draw[](n11) -- (n13);
\draw[](n12) -- (n14);
\draw[](n13) -- (n19);
\draw[](n14) -- (n16);
\draw[](n15) -- (n17);
\draw[](n15) -- (n16);
\draw[](n16) -- (n18);
\draw[](n17) -- (n20);
\draw[](n18) -- (n25);
\draw[](n19) -- (n21);
\draw[](n19) -- (n20);
\draw[](n20) -- (n22);
\draw[](n21) -- (n23);
\draw[](n22) -- (n24);
\draw[](n23) -- (n26);
\draw[](n23) -- (n24);
\draw[](n24) -- (n27);
\draw[](n25) -- (n34);
\draw[](n26) -- (n28);
\draw[](n27) -- (n29);
\draw[](n28) -- (n30);
\draw[](n29) -- (n31);
\draw[](n30) -- (n32);
\draw[](n31) -- (n33);
\draw[](n32) -- (n35);
\draw[](n32) -- (n33);
\draw[](n33) -- (n36);
\draw[](n34) -- (n37);
\draw[](n35) -- (n38);
\draw[](n36) -- (n39);
\draw[](n37) -- (n40);
\draw[](n38) -- (n41);
\draw[](n38) to[out=-50,in=50] (n40);
\draw[](n39) -- (n42);
\draw[](n39) -- (n40);
\draw[](n40) -- (n48);
\draw[](n41) -- (n43);
\draw[](n42) -- (n44);
\draw[](n43) -- (n45);
\draw[](n43) -- (n44);
\draw[](n44) -- (n46);
\draw[](n45) -- (n47);
\draw[](n46) -- (n51);
\draw[](n47) -- (n49);
\draw[](n48) -- (n50);
\draw[](n49) -- (n55);
\draw[](n50) -- (n52);
\draw[](n51) -- (n53);
\draw[](n51) -- (n52);
\draw[](n52) -- (n54);
\draw[](n53) -- (n56);
\draw[](n54) -- (n61);
\draw[](n55) -- (n57);
\draw[](n55) -- (n56);
\draw[](n56) -- (n58);
\draw[](n57) -- (n59);
\draw[](n58) -- (n60);
\draw[](n59) -- (n62);
\draw[](n59) -- (n60);
\draw[](n60) -- (n63);
\draw[](n61) -- (n70);
\draw[](n62) -- (n64);
\draw[](n63) -- (n65);
\draw[](n64) -- (n66);
\draw[](n65) -- (n67);
\draw[](n66) -- (n68);
\draw[](n67) -- (n69);
\draw[](n68) -- (n75);
\draw[](n69) -- (n71);
\draw[](n70) -- (n72);
\draw[](n71) -- (n73);
\draw[](n71) -- (n72);
\draw[](n72) -- (n74);
\draw[](n73) -- (n76);
\draw[](n74) -- (n80);
\draw[](n75) -- (n77);
\draw[](n75) -- (n76);
\draw[](n76) -- (n78);
\draw[](n77) -- (n79);
\draw[](n78) -- (n83);
\draw[](n79) -- (n81);
\draw[](n79) to[out=-50,in=50] (n80);
\draw[](n80) -- (n82);
\draw[](n81) -- (n87);
\draw[](n82) -- (n84);
\draw[](n83) -- (n85);
\draw[](n83) -- (n84);
\draw[](n84) -- (n86);
\draw[](n85) -- (n88);
\draw[](n86) -- (n93);
\draw[](n87) -- (n89);
\draw[](n87) -- (n88);
\draw[](n88) -- (n90);
\draw[](n89) -- (n92);
\draw[](n90) -- (n91);
\draw[](n92) -- (n94);
\draw[](n92) to[out=-50,in=50] (n93);
\draw[](n93) -- (n95);
\draw[](n94) -- (n96);
\draw[](n95) -- (n97);
\end{tikzpicture}
}
\end{minipage}
\begin{minipage}{0.140000\textwidth}
  \resizebox{\columnwidth}{1.2cm}{
\begin{tabular}{|c|c|}
\hline
\bfseries\footnotesize parameter &\bfseries\footnotesize angles\\
\hline
$\phi_{1}$& $25\pi/32$\\
$\phi_{2}$& $231\pi/256$\\
$\phi_{3}$& $97\pi/128$\\
$\phi_{4}$& $213\pi/512$\\
$\phi_{5}$& $847\pi/512$\\
$\phi_{6}$& $471\pi/256$\\
$\phi_{8}$& $719\pi/512$\\
\hline
\end{tabular}}
\end{minipage}
\begin{minipage}{0.110000\textwidth}
  \resizebox{\columnwidth}{1.2cm}{
\begin{tabular}{|c|c|}
\hline
$\phi_{9}$& $155\pi/128$\\
$\phi_{10}$& $915\pi/512$\\
$\phi_{11}$& $299\pi/512$\\
$\phi_{29}$& $3\pi/2$\\
$\phi_{30}$& $\pi$\\
$\phi_{32}$& $3\pi/4$\\
$\phi_{33}$& $\pi$\\
$\phi_{34}$& $\pi$\\
\hline
\end{tabular}}
\end{minipage}
\begin{minipage}{0.110000\textwidth}
  \resizebox{\columnwidth}{1.2cm}{
\begin{tabular}{|c|c|}
\hline
$\phi_{35}$& $39\pi/128$\\
$\phi_{36}$& $39\pi/128$\\
$\phi_{37}$& $\pi$\\
$\phi_{38}$& $\pi$\\
$\phi_{39}$& $\pi$\\
$\phi_{40}$& $217\pi/128$\\
$\phi_{41}$& $\pi$\\
$\phi_{42}$& $89\pi/128$\\
\hline
\end{tabular}}
\end{minipage}
\begin{minipage}{0.110000\textwidth}
  \resizebox{\columnwidth}{1.2cm}{
\begin{tabular}{|c|c|}
\hline
$\phi_{43}$& $7\pi/4$\\
$\phi_{45}$& $7\pi/4$\\
$\phi_{46}$& $3\pi/2$\\
$\phi_{47}$& $3\pi/2$\\
$\phi_{63}$& $129\pi/256$\\
$\phi_{66}$& $\pi$\\
$\phi_{68}$& $0$\\
$\phi_{69}$& $767\pi/512$\\
\hline
\end{tabular}}
\end{minipage}
\begin{minipage}{0.110000\textwidth}
  \resizebox{\columnwidth}{1cm}{
\begin{tabular}{|c|c|}
\hline
$\phi_{70}$& $535\pi/512$\\
$\phi_{71}$& $281\pi/512$\\
$\phi_{72}$& $991\pi/512$\\
$\phi_{73}$& $387\pi/512$\\
$\phi_{74}$& $613\pi/512$\\
$\phi_{75}$& $467\pi/512$\\
$\phi_{76}$& $\pi$\\
\hline
\end{tabular}}
\end{minipage}
\begin{minipage}{0.110000\textwidth}
  \resizebox{\columnwidth}{1cm}{
\begin{tabular}{|c|c|}
\hline
$\phi_{77}$& $255\pi/512$\\
$\phi_{78}$& $125\pi/256$\\
$\phi_{79}$& $1001\pi/512$\\
$\phi_{80}$& $\pi$\\
$\phi_{81}$& $\pi$\\
$\phi_{82}$& $\pi$\\
$\phi_{83}$& $45\pi/512$\\
\hline
\end{tabular}}
\end{minipage}
\begin{minipage}{0.110000\textwidth}
  \resizebox{\columnwidth}{1cm}{
\begin{tabular}{|c|c|}
\hline
$\phi_{84}$& $535\pi/512$\\
$\phi_{85}$& $495\pi/512$\\
$\phi_{86}$& $3\pi/2$\\
$\phi_{87}$& $479\pi/512$\\
$\phi_{88}$& $\pi$\\
$\phi_{89}$& $641\pi/512$\\
$\phi_{90}$& $497\pi/256$\\
\hline
\end{tabular}}
\end{minipage}
\begin{minipage}{0.110000\textwidth}
  \resizebox{\columnwidth}{1cm}{
\begin{tabular}{|c|c|}
\hline
$\phi_{91}$& $313\pi/256$\\
$\phi_{92}$& $643\pi/512$\\
$\phi_{93}$& $\pi$\\
$\phi_{94}$& $\pi$\\
$\phi_{95}$& $257\pi/256$\\
$\phi_{96}$& $53\pi/128$\\
$\phi_{97}$& $53\pi/128$\\
\hline
\end{tabular}}
\end{minipage}
\begin{minipage}{0.15\textwidth}
  \resizebox{\columnwidth}{3.3cm}{
\begin{tabular}{|c|c|}
\hline
\multicolumn{2}{|c|}{$\tau=0$}\\
\hline
$\phi_{{7}/{44}}$&$275\pi/256$\\
$\phi_{{12}/{48}}$&$217\pi/128$\\
$\phi_{{13}/{49}}$&$115\pi/128$\\
$\phi_{{14}/{50}}$&$249\pi/512$\\
$\phi_{{15}/{52}}$&$971\pi/512$\\
$\phi_{{16}/{51}}$&$33\pi/64$\\
$\phi_{{17}/{54}}$&$701\pi/512$\\
$\phi_{{18}/{53}}$&$973\pi/512$\\
$\phi_{{19}/{56}}$&$399\pi/256$\\
$\phi_{{20}/{55}}$&$79\pi/64$\\
$\phi_{{21}/{57}}$&$1007\pi/512$\\
$\phi_{{22}/{61}}$&$359\pi/256$\\
$\phi_{{23}/{59}}$&$963\pi/512$\\
$\phi_{{24}/{58}}$&$319\pi/256$\\
$\phi_{{25}/{64}}$&$211\pi/128$\\
$\phi_{{26}/{67}}$&$125\pi/256$\\
$\phi_{{27}/{62}}$&$617\pi/512$\\
$\phi_{{28}/{60}}$&$121\pi/256$\\
$\phi_{{31}/{65}}$&$719\pi/512$\\
\hline
\end{tabular}}
\end{minipage}
\begin{minipage}{0.15\textwidth}
\resizebox{\columnwidth}{3.3cm}{
\begin{tabular}{|c|c|}
\hline
\multicolumn{2}{|c|}{$\tau=1$}\\
\hline
$\phi_{{7}/{44}}$&$3\pi/2$\\
$\phi_{{12}/{48}}$&$\pi$\\
$\phi_{{13}/{49}}$&$755\pi/512$\\
$\phi_{{14}/{50}}$&$71\pi/128$\\
$\phi_{{15}/{52}}$&$863\pi/512$\\
$\phi_{{16}/{51}}$&$51\pi/32$\\
$\phi_{{17}/{54}}$&$491\pi/512$\\
$\phi_{{18}/{53}}$&$751\pi/512$\\
$\phi_{{19}/{56}}$&$777\pi/512$\\
$\phi_{{20}/{55}}$&$581\pi/512$\\
$\phi_{{21}/{57}}$&$205\pi/256$\\
$\phi_{{22}/{61}}$&$87\pi/128$\\
$\phi_{{23}/{59}}$&$759\pi/512$\\
$\phi_{{24}/{58}}$&$177\pi/512$\\
$\phi_{{25}/{64}}$&$505\pi/512$\\
$\phi_{{26}/{67}}$&$777\pi/512$\\
$\phi_{{27}/{62}}$&$\pi$\\
$\phi_{{28}/{60}}$&$317\pi/256$\\
$\phi_{{31}/{65}}$&$1015\pi/512$\\
\hline
\end{tabular}}
\end{minipage}
\begin{minipage}{0.15\textwidth}
\resizebox{\columnwidth}{3.3cm}{
\begin{tabular}{|c|c|}
\hline
\multicolumn{2}{|c|}{$\tau=2$}\\
\hline
$\phi_{{7}/{44}}$&$133\pi/256$\\
$\phi_{{12}/{48}}$&$21\pi/32$\\
$\phi_{{13}/{49}}$&$235\pi/512$\\
$\phi_{{14}/{50}}$&$527\pi/512$\\
$\phi_{{15}/{52}}$&$769\pi/512$\\
$\phi_{{16}/{51}}$&$769\pi/512$\\
$\phi_{{17}/{54}}$&$\pi$\\
$\phi_{{18}/{53}}$&$887\pi/512$\\
$\phi_{{19}/{56}}$&$51\pi/128$\\
$\phi_{{20}/{55}}$&$89\pi/128$\\
$\phi_{{21}/{57}}$&$455\pi/256$\\
$\phi_{{22}/{61}}$&$859\pi/512$\\
$\phi_{{23}/{59}}$&$931\pi/512$\\
$\phi_{{24}/{58}}$&$511\pi/256$\\
$\phi_{{25}/{64}}$&$31\pi/256$\\
$\phi_{{26}/{67}}$&$513\pi/512$\\
$\phi_{{27}/{62}}$&$151\pi/128$\\
$\phi_{{28}/{60}}$&$257\pi/512$\\
$\phi_{{31}/{65}}$&$175\pi/512$\\
\hline
\end{tabular}}
\end{minipage}
\begin{minipage}{0.15\textwidth}
\resizebox{\columnwidth}{3.3cm}{
\begin{tabular}{|c|c|}
\hline
\multicolumn{2}{|c|}{$\tau=3$}\\
\hline
$\phi_{{7}/{44}}$&$461\pi/512$\\
$\phi_{{12}/{48}}$&$\pi$\\
$\phi_{{13}/{49}}$&$253\pi/256$\\
$\phi_{{14}/{50}}$&$69\pi/128$\\
$\phi_{{15}/{52}}$&$279\pi/256$\\
$\phi_{{16}/{51}}$&$641\pi/512$\\
$\phi_{{17}/{54}}$&$1013\pi/512$\\
$\phi_{{18}/{53}}$&$7\pi/256$\\
$\phi_{{19}/{56}}$&$387\pi/256$\\
$\phi_{{20}/{55}}$&$959\pi/512$\\
$\phi_{{21}/{57}}$&$795\pi/512$\\
$\phi_{{22}/{61}}$&$249\pi/256$\\
$\phi_{{23}/{59}}$&$391\pi/256$\\
$\phi_{{24}/{58}}$&$267\pi/512$\\
$\phi_{{25}/{64}}$&$641\pi/512$\\
$\phi_{{26}/{67}}$&$59\pi/128$\\
$\phi_{{27}/{62}}$&$257\pi/512$\\
$\phi_{{28}/{60}}$&$181\pi/128$\\
$\phi_{{31}/{65}}$&$31\pi/64$\\
\hline
\end{tabular}}
\end{minipage}
\begin{minipage}{0.15\textwidth}
\resizebox{\columnwidth}{3.3cm}{
\begin{tabular}{|c|c|}
\hline
\multicolumn{2}{|c|}{$\tau=4$}\\
\hline
$\phi_{{7}/{44}}$&$31\pi/16$\\
$\phi_{{12}/{48}}$&$333\pi/512$\\
$\phi_{{13}/{49}}$&$97\pi/64$\\
$\phi_{{14}/{50}}$&$461\pi/512$\\
$\phi_{{15}/{52}}$&$347\pi/256$\\
$\phi_{{16}/{51}}$&$391\pi/256$\\
$\phi_{{17}/{54}}$&$479\pi/512$\\
$\phi_{{18}/{53}}$&$487\pi/512$\\
$\phi_{{19}/{56}}$&$23\pi/64$\\
$\phi_{{20}/{55}}$&$107\pi/64$\\
$\phi_{{21}/{57}}$&$51\pi/32$\\
$\phi_{{22}/{61}}$&$401\pi/256$\\
$\phi_{{23}/{59}}$&$397\pi/512$\\
$\phi_{{24}/{58}}$&$113\pi/64$\\
$\phi_{{25}/{64}}$&$47\pi/32$\\
$\phi_{{26}/{67}}$&$453\pi/512$\\
$\phi_{{27}/{62}}$&$927\pi/512$\\
$\phi_{{28}/{60}}$&$439\pi/512$\\
$\phi_{{31}/{65}}$&$89\pi/128$\\
\hline
\end{tabular}}
\end{minipage}
\begin{minipage}{0.15\textwidth}
\resizebox{\columnwidth}{3.3cm}{
\begin{tabular}{|c|c|}
\hline
\multicolumn{2}{|c|}{$\tau=5$}\\
\hline
$\phi_{{7}/{44}}$&$941\pi/512$\\
$\phi_{{12}/{48}}$&$419\pi/256$\\
$\phi_{{13}/{49}}$&$113\pi/64$\\
$\phi_{{14}/{50}}$&$79\pi/128$\\
$\phi_{{15}/{52}}$&$3\pi/2$\\
$\phi_{{16}/{51}}$&$257\pi/512$\\
$\phi_{{17}/{54}}$&$673\pi/512$\\
$\phi_{{18}/{53}}$&$73\pi/64$\\
$\phi_{{19}/{56}}$&$297\pi/256$\\
$\phi_{{20}/{55}}$&$667\pi/512$\\
$\phi_{{21}/{57}}$&$419\pi/512$\\
$\phi_{{22}/{61}}$&$97\pi/256$\\
$\phi_{{23}/{59}}$&$231\pi/256$\\
$\phi_{{24}/{58}}$&$87\pi/512$\\
$\phi_{{25}/{64}}$&$897\pi/512$\\
$\phi_{{26}/{67}}$&$513\pi/512$\\
$\phi_{{27}/{62}}$&$439\pi/512$\\
$\phi_{{28}/{60}}$&$0$\\
$\phi_{{31}/{65}}$&$211\pi/512$\\
\hline
\end{tabular}}
\end{minipage}

\caption{
  Our 3-qubit BEQS (see~\Cref{alg:exactgrover_blind}) for $w=012345$ and 
  $M_{POVM}=\{\ketbra0+\ketbra1,\ketbra2,\ketbra3,\ketbra4,\ketbra5\}$;
  thus, $N=6$ and $\tilde N=5$.
  Red nodes indicate blind oracles controlled by Oscar, 
  black nodes indicate Alice's computation.
  The measurement angles, which are specified to 10 bits, for each node
  is shown in the table; the
  measurement order is indicated with the node numbers.
  This computation has success probabilities $p_s\geq 1-10^{-4}$ for all
  queries $\tau\in w$.}
\label{fig:boc6}
\end{figure}

Unfortunately, \textsc{ObjPOVM} requires more resources than $f_p$ and
$f_b$ (\autoref{eq:objective_fpb}); therefore for reasons of economy, we set a
fixed oracle in every query, which results in fewer optimization
parameters.  One possible improvement is restricting  the legitimate
marked items $q\subset w$ to cut the loop at line~\ref{line:loop}
of~\cref{subr:obj_povm}.  Returning to the previous example where $w$=0123,
$M_{POVM}=\{\ketbra0+\ketbra1,\ketbra2+\ketbra3\}$, and the database entries are \{A,B\},
we simply set $q=\{0,2\}$. Whereas previously Oscar would randomly mark
item 0 or 1 to reveal A and would randomly mark 2 or
3 to reveal B, now Oscar marks only 0 to reveal A and marks 2 to reveal
B. This amount of speedup resulting from this strategy depends on how small $q$ compared to $w$. 

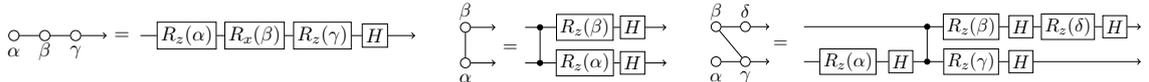
\begin{figure}[!h]
\begin{minipage}[!h]{.38\textwidth}
\resizebox{\columnwidth}{!}{
\begin{tikzpicture}
  [scale=.6,auto=left,every node/.style={inner sep=2pt,minimum size=5pt, draw=black}]
  \node[circle] (n1) at (0,0) {};
  \node[circle] (n2) at (1,0) {};
  \node[circle] (n3) at (2,0) {};
  \coordinate (n4) at (3,0);
  \node[draw=none] at ([yshift=-10pt]n1.south){$\alpha$};
  \node[draw=none] at ([yshift=-10pt]n2.south){$\beta$};
  \node[draw=none] at ([yshift=-10pt]n3.south){$\gamma$};
  \draw[->] (n3) -- (n4);
  \foreach \from/\to in {n1/n2,n2/n3} \draw (\from) -- (\to);
  \node[draw=none] (n5) at (3.5,0){$=$};
  \coordinate (n6) at (4,0);
  \coordinate (n7) at (4.1,0);
  \node[style={minimum size=13pt}](n8) at (5.6,0)    {$R_z(\alpha)$};
  \node[style={minimum size=13pt}](n9) at (7.8,0)  {$R_x(\beta)$};
  \node[style={minimum size=13pt}](n10) at (10,0) {$R_z(\gamma)$};
  \node[style={minimum size=13pt}](n11) at (11.7,0){$H$};
  \foreach \from/\to in {n7/n8,n8/n9,n9/n10,n10/n11} \draw (\from) -- (\to);
  \coordinate (n12) at (13,0);
  \draw[->] (n11.east) -- (n12);
\end{tikzpicture}
}
\end{minipage}
\hspace{.5em}
\begin{minipage}[!h]{0.2\textwidth}
\resizebox{\columnwidth}{!}{
\begin{tikzpicture}
  [scale=.6,auto=left,every node/.style={inner sep=2pt,minimum size=5pt, draw=black}]
  \node[circle] (n1) at (0,0) {};
  \node[circle] (n2) at (0,1.2) {};
  \coordinate (n3) at (1,0) {}; 
  \coordinate (n4) at (1,1.2) {};
  \node[draw=none] at ([yshift=-10pt]n1.south){$\alpha$};
  \node[draw=none] at ([yshift=10pt]n2.north){$\beta$};
  \draw (n1) -- (n2);
  \draw[->] (n1.east) -- (n3);
  \draw[->] (n2.east) -- (n4);
  \node[draw=none] at (1.5,0.5){$=$};

  \node[style={minimum size=13pt}](n5) at (4.0,0)   {$R_z(\alpha)$};
  \node[style={minimum size=13pt}](n6) at (4.0,1.2) {$R_z(\beta)$}; 
  \node[style={minimum size=13pt}](n7) at (5.6,0)   {$H$};
  \node[style={minimum size=13pt}](n8) at (5.6,1.2) {$H$};

  \node[circle,style={inner sep=0pt, minimum size=3pt, fill=black}](n9) at (2.5,0) {};
  \node[circle,style={inner sep=0pt, minimum size=3pt, fill=black}](n10) at (2.5,1.2) {};

  \coordinate (n00) at (2,0);
  \coordinate (n01) at (2,1.2);

  \foreach \from/\to in {n9/n10,n00/n9,n01/n10,n9/n5,n10/n6,n6/n8,n7/n5} \draw (\from) -- (\to);
\draw[->] (n7.east) -- ([xshift=20pt]n7.east);
\draw[->] (n8.east) -- ([xshift=20pt]n8.east);

\end{tikzpicture}
} 
\end{minipage}
\hspace{.5em}
\begin{minipage}[h!]{0.4\textwidth}
\resizebox{\columnwidth}{!}{
\begin{tikzpicture}[scale=.6,auto=left,every node/.style={inner sep=2pt,minimum size=5pt, draw=black}]
  \node[circle] (n1) at (0,0) {};
  \node[circle] (n2) at (0,1.2) {};
  \node[circle] (n3) at (1,0) {}; 
  \node[circle] (n4) at (1,1.2) {};
  \foreach \from/\to in {n1/n3,n2/n4,n2/n3} \draw (\from) -- (\to);
  \draw[->] (n3.east) -- ([xshift=18pt]n3.east);
  \draw[->] (n4.east) -- ([xshift=18pt]n4.east);
  \node[draw=none] at (2.2,0.6){$=$};

  \coordinate (n00) at (3,0);
  \coordinate (n01) at (3,1.2);
  \node[style={minimum size=13pt}](n5) at (4.5,0)        {$R_z(\alpha)$};
  \node[style={minimum size=13pt}](n6) at (8.7,1.2)    {$R_z(\beta)$};
  \node[style={minimum size=13pt}](n7) at (8.7,0)      {$R_z(\gamma)$};
  \node[style={minimum size=13pt}](n8) at (12,1.2)     {$R_z(\delta)$};
  \node[style={minimum size=13pt}](n9) at  (6.3,0)    {$H$}; 
  \node[style={minimum size=13pt}](n10) at (10.4,1.2) {$H$}; 
  \node[style={minimum size=13pt}](n11) at (10.4,0)   {$H$}; 
  \node[style={minimum size=13pt}](n12) at (13.6,1.2) {$H$}; 
  \node[draw=none] at ([yshift=-10pt]n1.south){$\alpha$};
  \node[draw=none] at ([yshift= 10pt]n2.north){$\beta$};
  \node[draw=none] at ([yshift=-10pt]n3.south){$\gamma$};
  \node[draw=none] at ([yshift= 10pt]n4.north){$\delta$};

  \node[circle,style={inner sep=0pt, minimum size=3pt, fill=black}](n13) at(7.2,0) {};
  \node[circle,style={inner sep=0pt, minimum size=3pt, fill=black}](n14) at(7.2,1.2) {};
  \draw(n13) -- (n14);

  \coordinate (n15) at (13.6,0);
  \foreach \from/\to in {n00/n5,n01/n6,n5/n9,n9/n7,n7/n11,n6/n10,n10/n8,n8/n12}  
  \draw (\from) -- (\to);
  \draw[->] (n11.east) -- ([xshift=25pt]n15.center);
  \draw[->] (n12.east) -- ([xshift=25pt]n12.center);

\end{tikzpicture}
} 
\end{minipage}
\hfill
\caption{Examples of computation within the gate model and the 1WQC
  model. The angles below nodes denote measurement angles; measurements are
  performed from the left to the right. The right hand side of each equation denotes
  the equivalent computation in the gate model, assuming all
  measurement outcomes zero.}
\label{fig:examples_mbqc}
\end{figure}

We test the BEQS for 3-qubit cases, obtaining quantum algorithms within
the gate model and the 1WQC model (this takes care of the BOQC model
also), where $w=012345$ and
$M_{POVM}=\{\ketbra{0}+\ketbra{1},\ketbra{2},\ketbra{3},\ketbra{4},\ketbra{5}\}$,
thus $N=6$ and  $\tilde N=5$. We choose this configuration based on its
potential to result in the smallest gate network based on the study of
\Cref{tab:CNOT_counting}.  We obtain \Cref{fig:circuit6} for the gate
model, that is,  a circuit comprising \{CNOT, $U$\}, where $U$ has a
form of~\autoref{eq:u}.  For the BOQC result, we obtain \Cref{fig:boc6}.
The cluster state is obtained by optimizing networks comprising
\{CPHASE, $R_z$\}, where $R_z(\alpha)\equiv e^{\frac{-i\alpha Z}{2}}$,
then transform the result into a graph state, whose measurement angles,
along with the $\alpha$ parameters, are the optimization parameters.
See~\Cref{fig:examples_mbqc} for some examples of the corresondence
between the gate model and the 1WQC model.  In our optimization, we set
$\delta=10^{-4}$ for both models, resulting in precisions
$\varepsilon<2.1\times 10^{-9}$ for the gate model and
$\varepsilon<1.2\times 10^{-10}$ for the BOQC model.

\begin{figure}[!h]
  \centering
  \begin{minipage}{0.2\textwidth}
    \begin{tikzpicture}
    [font=\tiny,scale=.55,auto,every node/.style={circle,inner sep=0pt,minimum size=7.8pt,draw=black,text width=7.9pt,align=center}]
\node[] (n1) at (0,1){2};
\node[] (n2) at (0,0){1};
\node[red, color=red] (n3) at (1,1){3};
\node[red, color=red] (n4) at (1,0){4};
\node[red, color=red] (n5) at (2,1){5};
\node[red, color=red] (n6) at (2,0){6};
\node[] (n7) at (3,1){7};
\node[] (n8) at (3,0){8};
\node[] (n9) at (4,1){10};
\node[] (n10) at (4,0){9};
\draw[](n1) -- (n3);
\draw[](n2) -- (n4);
\draw[](n3) -- (n5);
\draw[](n3) -- (n4);
\draw[](n4) -- (n6);
\draw[](n5) -- (n7);
\draw[](n6) -- (n8);
\draw[](n7) -- (n9);
\draw[](n8) -- (n10);
\draw[](n9) -- (n10);
    \end{tikzpicture}
\end{minipage}
\begin{minipage}{0.140000\textwidth}
  \resizebox{\columnwidth}{.6cm}{
\begin{tabular}{|c|c|}
\hline
\bfseries\footnotesize parameter &\bfseries\footnotesize angle\\
\hline
$\phi_{1}$& $0$\\
$\phi_{2}$& $0$\\
\hline
\end{tabular}}
\end{minipage}
\begin{minipage}{0.08000\textwidth}
  \resizebox{\columnwidth}{.6cm}{
\begin{tabular}{|c|c|}
\hline
$\phi_{3}$& $0$\\
$\phi_{4}$& $0$\\
$\phi_{7}$& $0$\\
\hline
\end{tabular}
}
\end{minipage}
\begin{minipage}{0.08000\textwidth}
  \resizebox{\columnwidth}{.6cm}{
\begin{tabular}{|c|c|}
\hline
$\phi_{8}$& $0$\\
$\phi_{9}$&  $\pi$\\
$\phi_{10}$& $\pi$\\
\hline
\end{tabular}
}
\end{minipage}
\begin{minipage}{0.08000\textwidth}
  \resizebox{\columnwidth}{.6cm}{
\begin{tabular}{|c|c|}
\hline
\multicolumn{2}{|c|}{$\tau=0$}\\
\hline
$\phi_{5}$& $\pi$\\
$\phi_{6}$& $\pi$\\
\hline
\end{tabular}}
\end{minipage}
\begin{minipage}{0.08000\textwidth}
  \resizebox{\columnwidth}{.6cm}{
\begin{tabular}{|c|c|}
\hline
\multicolumn{2}{|c|}{$\tau=1$}\\
\hline
$\phi_{5}$& $\pi$\\
$\phi_{6}$& $0$\\
\hline
\end{tabular}}
\end{minipage}
\begin{minipage}{0.08\textwidth}
  \resizebox{\columnwidth}{.6cm}{
  \begin{tabular}{|c|c|}
\hline
\multicolumn{2}{|c|}{$\tau=2$}\\
\hline
$\phi_{5}$& $0$\\
$\phi_{6}$& $\pi$\\
\hline
\end{tabular}}
\end{minipage}
\begin{minipage}{0.08000\textwidth}
  \resizebox{\columnwidth}{.6cm}{
\begin{tabular}{|c|c|}
\hline
\multicolumn{2}{|c|}{$\tau=3$}\\
\hline
$\phi_{5}$& $0$\\
$\phi_{6}$& $0$\\
\hline
\end{tabular}}
\end{minipage}
\caption{The 2-qubit Grover algorithm within the BOQC scheme, with $N=\tilde N=4$,
$w=0123$, and $M_{POVM}=\{\ketbra0,\ketbra1,\ketbra2,\ketbra3\}$;
for notations, follow~\autoref{fig:boc6}.} 
\label{fig:boc4}
\end{figure}
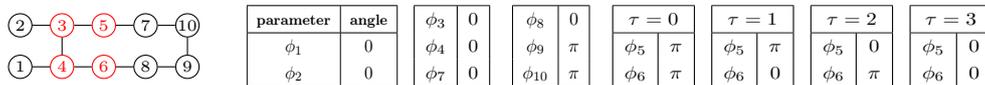

We have demonstrated that BEQS obtains exact quantum search algorithms
with blind oracles for two computation models. Moreover, BEQS has
reduced the size of computation for a five-entry database ($\tilde N=5$)
from using 19 CNOT gates (see~\autoref{tab:CNOT_counting}) to 17 CNOT
gates (see~\autoref{fig:circuit6}).  Our work establishes the
unfortunate fact that the implementation complexity grows very rapidly
for the Grover algorithm in the BOQC model.  As a comparison, we obtain
a cluster state for the 2-qubit Grover algorithm in~\autoref{fig:boc4},
where $\tilde N=4$ and $w=0123$.  Going from a four-element database to
a five-element database for an exact quantum search algorithm within the
BOQC scheme, means going from a 10-node to a 97-node cluster state.

\subsection{BOQC on NV-centers: the implementation of BEQS algorithms}
\label{sec:nvc}
Here we introduce our proposal to implement a BOQC computation using
NV-centers. We propose a direct realization of the results shown above:
a physical implementation of 3 qubit BEQS (\autoref{fig:boc6}), and of
the 2-qubit Grover algorithm (\autoref{fig:boc4}).  The main
challenges for physical implementation are the sizable physical
resources --- we need 97 qubits to run 3-qubit BEQS --- and the high
fidelity transmission of encrypted qubits from Alice or Oscar to Bob.
We think that these challenges can be at least largely overcome:  To
deal with the large size, we note the possibility of ``reusing'' the
qubits~\cite{housmand2018}.  To accomplish reliable transmission, we
propose using remote state preparation (RSP)~\cite{bennett2001remote} as
a quantum channel.  The re-use strategy drastically decreases the number
of qubits: from 97 to 4 qubits for 3-qubit BEQS and from 10 to 3 qubits
for 2-qubit Grover.  Moreover, RSP is understood to be very efficient
for the family of states to be transmitted~\cite{lo2000classical}; for
RSP in our setting no additional classical communication at all is
needed, automatically maintaining the blindness of the scheme.  

\begin{figure}[!t]
  \begin{minipage}[!b]{0.8\textwidth}
  \begin{tikzpicture}[>=stealth,scale=1,auto=left, style={inner sep=2pt}]
    \node (alice) at (0,1) {\includegraphics[width=.8cm]{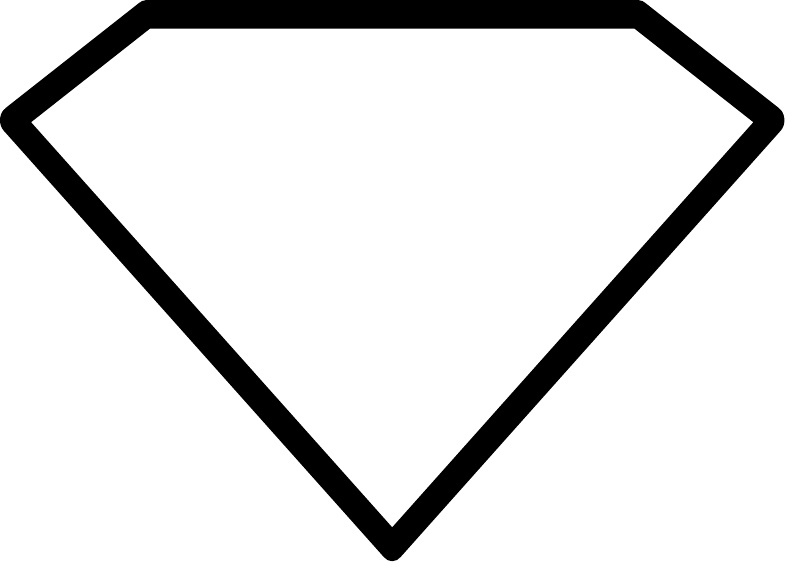}};
    \node (oscar) at (8,1) {\includegraphics[width=.8cm]{diamond}};
    \node (bob)   at (4,0) {\includegraphics[width=2.3cm]{diamond}};

    \node [circle, draw=black, inner sep=0.5pt]  at (2,1) {\tiny 1};
    \node [circle, draw=black, inner sep=0.5pt]  at (6,1) {\tiny 1};
    \node [circle, draw=black, inner sep=0.5pt]  at (5.4,-.25) {\tiny 2};
    \node [circle, draw=black, inner sep=0.5pt]  at (4, -0.45){\tiny 3};
    \node [circle, draw=black, inner sep=0.5pt]  at (6.45, -0.25){\tiny 4};

    \node (Delta) at (.8,1.6) {$\Delta_j$};
    \node (delta) at (7.2,1.6) {$\delta_j$};

    \node at ([yshift=1em]alice.north) {Alice};
    \node at ([yshift=1em]oscar.north) {Oscar};
    \node at ([yshift=1em]bob.north) {Bob};

    \node [circle,style={fill=black,inner sep=1.3pt}] (ea) at (alice.center){};
    \node [circle,style={fill=black,inner sep=1.3pt}] (eo) at (oscar.center){};
    \node [circle,style={fill=black,inner sep=1.3pt}](eb) at ([yshift=10pt]bob.center){};
    \coordinate (e1) at ([xshift=-15pt]eb.center){};
    \coordinate (e2) at ([xshift=-5pt]eb.center){};
    \coordinate (e3) at ([xshift=5pt]eb.center){};
    \coordinate (e4) at ([xshift=15pt]eb.center){};

    \node [circle,style={fill=red}](n1) at ([yshift=-7pt]e1.south){};
    \node [circle,style={fill=green}](n2) at ([yshift=-7pt]e2.south){};
    \node [circle,style={fill=blue}](n3) at ([yshift=-7pt]e3.south){};
    \node [circle,style={fill=yellow}](n4) at ([yshift=-7pt]e4.south){};

    \path[draw,style={decorate,decoration={snake},sloped}] (ea) -- (eb);
    \path[draw,style={decorate,decoration={snake},sloped}] (eo) -- (eb);

\node[cloud, cloud puffs=10, cloud ignores aspect, minimum width=1cm,
minimum height=1cm, align=center, draw] (sea)  at ([xshift=-1.3cm]ea)
{\tiny $\ket*{+_{\Theta_j}}$};

    \path[draw] (sea) -- (ea);

\node[cloud, cloud puffs=10, cloud ignores aspect, minimum width=1cm,
minimum height=1cm, align=center, draw] (seo)  at ([xshift=1.3cm]eo)
{\tiny $\ket*{+_{\theta_j}}$};
    \path[draw] (seo) -- (eo);

    \node (swapc) at ([xshift=25pt]n4) {\small$\times$};  
    \node (swapt) at ([xshift=25pt]e4) {\small $\times$};
    \path [draw] (swapc.center) -- ([xshift=-8pt]swapc.center);
    \path [draw] (swapt.center) -- ([xshift=-8pt]swapt.center);
    \path [draw] (swapc.center) -- ([xshift=8pt]swapc.center);
    \path [draw] (swapt.center) -- ([xshift=8pt]swapt.center);
    \path [draw] (swapt.center) -- (swapc.center);
    
    \node at([xshift=11pt]swapc.center){,};
    \node at([xshift=25pt]swapc.center){\resizebox{2em}{!}{$\Qcircuit@C=1em@R=1em{&\measureD{+_{\delta_j}}}$}};

    \node[circle,style={fill=black,inner sep=.8pt}] (phc) at ([yshift=-8pt]n2) {};  
    \node[circle,style={fill=black,inner sep=.8pt}] (pht) at ([yshift=-8pt]n3) {};
    \path [draw] (phc.center) -- ([yshift=6pt]phc.center);
    \path [draw] (pht.center) -- ([yshift=6pt]pht.center);
    \path [draw] (phc.center) -- ([yshift=-6pt] phc.center);
    \path [draw] (pht.center) -- ([yshift=-6pt] pht.center);
    \path [draw] (pht.center) -- (phc.center);
  \end{tikzpicture}
\end{minipage}
\hfill
\begin{minipage}[!b]{0.16\textwidth}
  \resizebox{\columnwidth}{!}{
  \begin{tikzpicture}
[font=\tiny,scale=.5,auto,every node/.style={circle,inner sep=0pt,minimum size=7.4pt,draw=black,text width=7.5pt,align=center}]
\node[label={[shift={(0.25,-0.5)},color=gray]1},draw=red, fill=red, text=white] (n1) at (0,1){2};
\node[label={[shift={(0.25,-0.5)},color=gray]1},draw=red, fill=red, text=white] (n2) at (0,0){1};
\node[label={[shift={(0.25,-0.5)},color=gray]2},draw=blue, fill=blue, text=white] (n3) at (1,1){3};
\node[label={[shift={(0.25,-0.5)},color=gray]2},draw=green, fill=green] (n4) at (1,0){4};
\node[label={[shift={(0.25,-0.5)},color=gray]3},draw=red, fill=red, text=white] (n5) at (2,1){5};
\node[label={[shift={(0.25,-0.5)},color=gray]3},draw=blue, fill=blue, text=white] (n6) at (2,0){6};
\node[label={[shift={(0.25,-0.5)},color=gray]4},draw=green, fill=green] (n7) at (3,1){7};
\node[label={[shift={(0.25,-0.5)},color=gray]4},draw=red, fill=red, text=white] (n8) at (3,0){8};
\node[label={[shift={(0.25,-0.5)},color=gray]5},draw=blue, fill=blue, text=white] (n9) at (4,1){10};
\node[label={[shift={(0.25,-0.5)},color=gray]5},draw=green, fill=green] (n10) at (4,0){9};
\draw[](n1) -- (n3);
\draw[](n2) -- (n4);
\draw[](n3) -- (n5);
\draw[](n3) -- (n4);
\draw[](n4) -- (n6);
\draw[](n5) -- (n7);
\draw[](n6) -- (n8);
\draw[](n7) -- (n9);
\draw[](n8) -- (n10);
\draw[](n9) -- (n10);
\begin{scope}[decoration={markings,mark=at position 0.6 with {\arrow{>}}}]
\draw[postaction={decorate}](n1) -- (n3);
\draw[postaction={decorate}](n2) -- (n4);
\draw[postaction={decorate}](n3) -- (n5);
\draw[postaction={decorate}](n4) -- (n6);
\draw[postaction={decorate}](n5) -- (n7);
\draw[postaction={decorate}](n6) -- (n8);
\draw[postaction={decorate}](n7) -- (n9);
\draw[postaction={decorate}](n8) -- (n10);
\end{scope}
\begin{scope}[every node/.style={draw=none,align=center,inner sep=1pt}]
  \node[label=below:$\mathcal A$] (a1) at (0,-1) {};
  \node[label=below:$\tilde{\mathcal O}$] (o1) at (1.5,-1) {};
  \node[label=below:$\mathcal D_{\pi}$] (d1) at (3.5,-1) {};
  \draw[-] (0,-.8)  -- (0,-.5);
  \draw[|-|] (1,-.7) -- (2,-.7) ;
  \draw[|-|] (3,-.7) -- (4,-.7) ;
\end{scope}

\end{tikzpicture}
}
\end{minipage}
  \begin{minipage}[!t]{\textwidth}
    \resizebox{\columnwidth}{!}{
     \input{boc_colored}
    }
\end{minipage}

  \caption{ 
      Optimized protocol using
      three NV-center nodes for the implementation of BOQC.
      The 97-node graph is the 3-qubit BEQS
      computation and the 10-node graph is the 2-qubit Grover
      algorithm.  Within the diamonds, the black node represents an electron
      spin and other colors represent nuclear spins; the node in the
      graph state will be assigned to the nuclear spin that has the same
      color.  Bob uses his electron spin for multiple purposes: 1) as an interface to create a
      quantum channel with the clients, 2) as a medium to perform CPHASE gates
      between nuclear spins, and 3) as an ancilla to measure his nuclear
      spins.  Alice and Oscar alternately control the computation,
      which comprises $\abs{\mathcal{FG}}$  rounds of steps \circled{1}--\circled{4} with
      the ordering shown as node numbers, where $\mathcal{FG}$ is the
      total graph.  The following sequence
      describes a round of Alice's moves.~\circled{1} RSP: Alice and Bob
      perform a heralded
      entanglement~\cite{barrett2005efficient,bernien2013heralded} to
      share a singlet state
      $\ket{\Psi^-}=\frac{\ket{01}-\ket{10}}{\sqrt2}$ between their two
      electron spins. Alice has in mind that she wishes to deliver state $\ket{+_{\Theta_j}}$ to Bob.  She accomplishes this by measuring her electron spin (i.e., her half of the singlet state) in $\Theta_j$;
      depending on her measurement outcome $s_j$, Bob receives
      $\frac{\ket0+e^{i(\Theta_j+s_j\pi)}\ket 1}{\sqrt2}$, where he knows neither $\Theta_j$ nor $s_j$.
      \circled{2} Bob swaps the electron spin with the nuclear spin
      according to the color.
      Bob buffers all the required qubits by repeating
      steps~\circled{1}--\circled{2}, which can also be from
      Oscar.~\circled{3} Bob applies CPHASE gates --- connecting the
      nodes in the graph --- 
      according to the subgraph; he connects only the nearest
      neighbor of the node that he is going to measure. 
      ~\circled{4} Bob measures qubit $j$ in $\Delta_j$ --- as
      instructed by Alice --- then announces
      measurement outcome $b_j$ using a public broadcast channel. The angle
      $\Delta_j$ is computed by Alice taking account all the
      corrections. The same procedure applies to Oscar for his
      computations.  This process is repeated until all nodes are
      measured.  It is worth mentioning that the grey numbers represent a
      partial ordering induced by flow, which was computed using an
      algorithm of~\cite{mhalla2008finding}.  Indeed, the total ordering
      may be selected arbitrarily as long as the partial ordering is
      respected. 
  } \label{fig:nvc} \end{figure}



For the BOQC implementations that we propose, the total graphs together
with measurement angles are shown in~\autoref{fig:boc6} for 3-qubit BEQS
and \autoref{fig:boc4} for the 2-qubit Grover algorithm.  Given that
Alice and Oscar have successfully shared the key $\vec\xi$, and all
parties agreed upon a total ordering and graphs, they thus run the
protocol shown in~\autoref{fig:nvc}.  In this implementation, the BOQC
protocol (see~\autoref{fig:boqc_protocol}) is optimized to perform the
computations per partition.  Moreover, every partition is divided into
smaller subgraphs  with size four (three for 2-qubit Grover).  This
strategy assumes that measured qubits can reset and reused.  Processing
a computation of a subgraph is described as
steps~\circled{1}--\circled{4} in~\autoref{fig:nvc}.  Compared to the 
UBQC scheme, an additional set of corrections $\vec s$ appears as a result from
the RSP. 

Based on state-of-the-art
technologies~\cite{humphreys2018deterministic,kalb2017entanglement,cramer2016repeated},
we estimate the total computation time for 3-qubit BEQS to be 3 seconds,
where the time for each step is $t_1\approx 25~\text{ms}, t_2\approx
1.5~\mu\text{s}, t_3\approx 3.5~\text{ms}, t_4~\approx 0.5~\text{ms}$;
assuming every two-bit operation requires 0.5 ms and all the
operations on the nuclear spins are performed by coupling the electron
spin. For instance, given nuclear spins $\{n_1,n_2\}$ and electron
spin $e$, one applies the CPHASE gate between $n_1$ and $n_2$ as follows:
SWAP($n_1,e$)-CPHASE($e,n_2$)-SWAP($n_1,e$), where each SWAP gate is
implemented with three CNOT gates. For the 2-qubit Grover algorithm,
the total run time is estimated to be 305 ms.

Processing one subgraph, that is, executing
steps~\circled{1}--\circled{4}, requires 1--2 heralded entanglements and
1--2 CPHASE operations, which, on average, is completed within 31 ms for
both computations.  After measuring a qubit, some qubits are idle until
being measured.  For the 3-qubit BEQS, the idling qubits need to
maintain their coherence for around 91 ms, while around 3 RSPs and other
operations are performed on the electron spin.  The worst idle case
happens at node 23, with idle time 370 ms while 11 RSPs are performed on
the electron spin. For the 2-qubit Grover, the idle average is 54 ms,
while around 2 RSPs and other operations are performed on the electron
spin; the worst idle happens at node 4, which is 170 ms while 5 RSPs are
performed on the electron spin.  We observe that the ordering we have
shown here is not the optimum strategy from the point of view of the
idling.  It would be possible to insert some redundant nodes to reduce
the idle time.

Nuclear spins on NV-centers have been reported to possess coherence time
of more than 1 second, even at room
temperature~\cite{maurer2012room,dolde2013room}. Moreover, high-fidelity
one- and two-bit gates operation on the nuclear spins have been
demonstrated~\cite{taminiau2014universal}. While the coherence time
seems sufficient, the activities on the electron spins can decohere the
idle nuclear spins, especially during the heralded entanglement
attempts. Moreover, our time estimation uses the best rate known of
heralded entanglement, with the nodes separated by 2
meters~\cite{humphreys2018deterministic}.  However, the obtained
herlded-entanglement fidelity is still low, hence a distillation scheme would needed.  We
conclude that the current technology is still not yet ready to implement the
algorithms in~\autoref{fig:nvc}, but foreseeable improvements would make it possible.

\section*{Acknowledgment}
We thank Tim Taminiau and  Slava Dobrovitski for the insightful
discussions about NV-centers, to Tal Mor and Rotem Liss for the discussions
about distributed and blind computations, to Barbara Terhals's group in
Delft and to IQI members in Aachen for various discussions. We
acknowledge the support of Forschungzentrum Jülich for the access to
JURECA and RWTH Aachen for the access to clusters.

\printbibliography

\appendix
  \section{Appendix}

\subsection{The arbitrary step of search iteration}
\label{subsection:thearbitrary}

This section supplements \autoref{sec:exact_grover-hoyer}, namely
finding the angles $\psi,\varphi$ of the H\o{}yer amplitude
amplification within the operator
$Q(\psi,\varphi)$~\cite{hoyer2000arbitrary}.  Whereas in the Grover
algorithm one iteration is restricted to the rotation by $2\theta_0$,
the H\o{}yer amplitude amplification allows a rotation within the range
$[-2\theta_0,2\theta_0]$, where $\theta_0$ is the initial angle. 

Suppose that we employ $n$ qubits and start with an equal superposition
of $N$ basis states $\ket*{\Psi_{init}}$ where $2^{n-1}\leq N\leq 2^n$. Let
$x$ be the indices that can be realized by $n$ qubits,
$x=\{0,\dots,2^n-1\}$ and $W$ be a set of all possible subsets of
the $N$-element database, $D=\{w\subseteq x : \norm{w}=N \}$, and let $w\in
W$, then 
\begin{equation}
\ket*{\Psi_{init}}=\frac{1}{\sqrt{N}}\sum_{j\in w}\ket*{j}.  
\end{equation}

Assume that we have an oracle that implements some function $f$ that can
distinguish weather a state is the target. Let $y$ be the set of targets,
the action of $f$ be
\begin{equation}
f(j)=
\begin{cases}
    1, \quad  \text{if}\,\,j\in y   \\
    0, \quad  \text{if}\,\,j\in x\setminus y. 
    \end{cases}
\end{equation}

The function $f$ induces a subspace spanned by ``good state''
$\ket*{\Psi_1}=\frac{1}{\sqrt{N}}\sum_{\{j: f(j)=1\}}\ket*{j}$ and ``bad
state'' $\ket*{\Psi_0}=\frac{1}{\sqrt{N}}\sum_{\{j: f(j)=0\}}\ket*{j}$.
Thus, the initial state can be rewritten as
$\ket*{\Psi_{init}}=\ket*{\Psi_1}+\ket*{\Psi_0}$. Let us search for $M$
targets. In the normalized basis of good and bad states, we rewrite  
again the initial state  
\begin{equation}
  \ket*{\Psi_{init}}=\sqrt{a}\ket*{\tilde\Psi_1}+
\sqrt{1-a}\ket*{\tilde\Psi_0}
\end{equation}
where
$\ket*{\tilde\Psi_1}=\frac{1}{\sqrt{M}}\ket*{\Psi_1}$, 
$\ket*{\tilde\Psi_0}=\frac{1}{\sqrt{N-M}}\ket*{\Psi_0}$, and 
$a=\frac{M}{N} \equiv \sin^2(\theta_0)$. 

Let $Q(\varphi,\psi)$ be the operator that performs search
iteration with parameters $\varphi,\psi\in [0,2\pi)$ 
\begin{equation}
  Q(\varphi,\psi)\equiv -\mathcal A S_0(\psi)\mathcal A
  S_y(\varphi).
\end{equation}
Where $\mathcal A$ is the preparation operator that transforms state
$\ket*{0}^{\otimes n}$ into the equal superposition state $\mathcal
A\ket*{0}^{\otimes n}=\ket*{\Psi_{init}}$. In the Grover algorithm of database size $2^n$,
$\mathcal A$ basically consists of Hadamards. Note that we can prepare
$\ket*{\Psi_{init}}$ from any convenient starting state. For simplicity, we start
with zero state $\ket*{0}^{\otimes n}$.  

Essentially, $Q(\varphi,\psi)$ consists of one oracle call
$S_y(\varphi)$ and a
diffusion operator $D(\psi)\equiv\mathcal A S_0(\psi)\mathcal A$.
The oracle call $\mathcal S_y(\varphi)$ ``marks'' the
targets $y$ by $e^{i\varphi}$ and it can be defined as  
\begin{equation}
  S_y(\varphi)\coloneqq I-(1-e^{i\varphi})\ket*{\tilde\Psi_1}\bra*{\tilde\Psi_1}.
\end{equation} 
The operator $S_0(\psi)$ marks the state before preparation (in our case
was $\ket*{0}^{\otimes n}$) with phase $e^{i\psi}$. Thus, the diffusion
operator follows 
\begin{equation}
  D(\psi)
  = \mathcal A \left[I - (1-e^{i\psi})(\ket*{0}\bra{0})^{\otimes n}\right] \mathcal A 
  = I - (1-e^{i\psi}) \ket*{\Psi_{init}}\bra{\Psi_{init}}. 
\end{equation}

By using basis $\{\ket*{\tilde\Psi_0},\ket*{\tilde\Psi_1}\}$, we can
represent $Q$ in matrix form 
\begin{equation}
  \label{eqn:qmatrix}
  Q(\varphi,\psi) = 
  \begin{bmatrix}
    -a(1-e^{i\psi})-e^{i\psi}  & (1-e^{i\psi})e^{i\varphi} \sqrt{a(1-a)} \\
    (1-e^{i\psi})\sqrt{a(1-a)} &
    a(1-e^{i\psi})e^{i\varphi}-e^{i\varphi}
  \end{bmatrix}.
\end{equation}

Now, the question is: how to implement an arbitrary rotation $\theta$
from $\ket*{\Psi_{init}}$ by applying $Q(\varphi,\psi)$? We need to find
out what are $\varphi$ and $\psi$ given $\theta$.  By imposing
some conditions on $\varphi$ and $\psi$, we can find them by using some
tricks.

Note that we are only working with two dimensional Hilbert space 
spanned by the complex vectors
$\{\ket*{\tilde\Psi_0},\ket*{\tilde\Psi_1}\}$. Therefore, we may associate
$Q$ with some general form of two-dimensional unitary operator. 

Given an arbitrary unitary operator $U$ with four parameters
$\delta,\varphi_1,\varphi_2,\theta \in [0,2\pi)$
\begin{equation} 
  U = e^{i\frac{\delta}{2}}
  \begin{pmatrix}
    e^{i\varphi_1}\cos(\theta) & e^{i\varphi_2}\sin(\theta)\\
  -e^{i\varphi_2}\sin(\theta) & e^{-i\varphi_1}\cos(\theta)
  \end{pmatrix}.
\end{equation}
Let us transform the parameters into the following. Let
$\varphi_1=\mu+\nu$ and $\varphi_2=\mu-\nu+\pi$, thus 
\begin{equation} 
  U = e^{i\frac{\delta}{2}}
  \begin{pmatrix}
    e^{i\mu+i\nu}\cos(\theta) & -e^{i\mu-i\nu}\sin(\theta)\\
   e^{i\mu-i\nu}\sin(\theta) & e^{-i\mu-i\nu}\cos(\theta)
  \end{pmatrix}.
\end{equation}

We impose the condition that the diagonal elements be equal;
that is fulfilled if and only if $\varphi_1=-\varphi_1$. This implies
$\varphi_1=0$ and thus $\nu=-\mu$. Let us call this matrix
$\tilde U$. Latter, we re-parameterize $\tilde U$ by setting
$\delta/2=v$ and $2\mu=u$, thus
\begin{equation} 
  \tilde U = 
  e^{i\frac{\delta}{2}}
  \begin{pmatrix}
     \cos(\theta) & -e^{i2\mu}\sin(\theta)\\
   e^{-i2\mu}\sin(\theta) & \cos(\theta)
 \end{pmatrix}
 =
  e^{iv}
  \begin{pmatrix}
     \cos(\theta) & -e^{iu}\sin(\theta)\\
     e^{-iu}\sin(\theta) & \cos(\theta)
 \end{pmatrix}.
\end{equation}

We factorize $\tilde U$ in the following way:
\begin{equation}
  \tilde U
 = e^{iv}
  \begin{pmatrix}
    {1} & 0 \\
    0        & e^{-iu}
  \end{pmatrix}
  \begin{pmatrix}
     \cos(\theta) & -\sin(\theta)\\
     \sin(\theta) & \cos(\theta)
  \end{pmatrix}
  \begin{pmatrix}
    1   & 0 \\
    0  & e^{iu}
  \end{pmatrix}.
\end{equation}
In this form, it is easy to see that $\tilde U$ performs a real rotation up to
some conditional phases.  
The aim is to associate our search operator $Q$ with $\tilde U$.  

We set $Q$ such that its diagonal elements are also equal, which means 
$
  -a(1-e^{i\psi})-e^{i\psi} = a(1-e^{i\psi})e^{i\varphi}-e^{i\varphi}.
$
From the H\o yer's result~\cite{hoyer2000arbitrary}, suppose
$\varphi\neq\pi$, then this condition is fulfilled if and only if 
\begin{equation}
  \tan (\varphi/2) = \tan(\psi/2)(1-2a).
  \label{eqn:HoyerRes}
\end{equation}

Given $\tilde Q$, which is the matrix $Q$ with equal diagonal elements.
At this point, it is straightforward to parameterize $\tilde Q$.  We can
find parameters $\psi$ and $\varphi$ in the following manner 
\begin{align}
  \norm{(1-e^{i\psi})\sqrt{a(1-a)}} &=\sin(\theta) \nonumber \\
  \psi &= \arccos\left(1-\frac{\sin^2(\theta)}{2a(1-a)}\right) \\ 
  \varphi &= 2\arctan(\psi/2)(1-2a).
  \label{eqn:grover_psi_and_phi}
\end{align}

Now we are able to perform an arbitrary rotation $\theta$ on a state
$\ket*{\Psi_{init}}$ with initial angle $\theta_0=\arcsin(\sqrt{a})$
using $\tilde Q(\varphi,\psi)$, up to some conditional phases.  Thus, we
may relate $Q$ and $\tilde Q$ by canceling its conditional phases  
\begin{equation}
  Q=e^{-iv}
  \begin{pmatrix}
    {1} & 0 \\
    0        & e^{iu}
  \end{pmatrix}
  \tilde Q
  \begin{pmatrix}
    {1} & 0 \\
    0        & e^{-iu}
  \end{pmatrix}.
\end{equation}

Two additional parameters are necessary in order to have a correct
rotation, thus $Q=Q(\varphi,\psi,u,v)$. By knowing $\psi$, the phases $u$ and
$v$ can be obtained straightforwardly, for instance
\begin{align}
  v &= \arg\left(-a(1-e^{i\psi})-e^{i\psi}\right)  \\
  u &= v - \arg\left((1-e^{i\psi})\sqrt{a(1-a)}\right).
\end{align}

Since one rotation is limited to
{$\theta\in[-2\theta_0,2\theta_0]$}, we need to split it
into several iterations if $\norm{\theta}>\norm{2\theta}$. Suppose we
perform $m>1$ iterations for which each iteration rotates 
$\tilde\theta=\theta/m$ with parameters $\tilde u,\tilde v$, thus  
\begin{equation}
  Q^m=e^{-i m\tilde v}
  \begin{pmatrix}
    {1} & 0 \\
    0   & e^{i \tilde u}
  \end{pmatrix}
  \begin{pmatrix}
    \cos(\tilde\theta)&-\sin(\tilde\theta)\\
    \sin(\tilde\theta)&\cos(\tilde\theta)
  \end{pmatrix}^m
  \begin{pmatrix}
    {1} & 0 \\
    0        & e^{-i\tilde u}
  \end{pmatrix}.
\end{equation}
For the identical iterations, the phase corrections need only be performed
once at the beginning and end, since they will be canceled out in the intermediate stages.

  \subsection{The exhaustive search circuit}
\label{sec:results}
\begin{minipage}{\linewidth}
\bigskip
\centering
\begin{tabular}{|C{1.3cm}|p{13cm}|} \hline
 \textbf{N} & 
\multicolumn{1}{>{\centering\arraybackslash}m{13cm}|}{\textbf{Equivalent
combinations of database}} \\ \hline 
& 
01234, 01235, 01236, 01237, 01245, 01246, 01345, 01357,
01456, 01457, 02346, 02367, 02456, 02467, 04567, 12357,
12367, 13457, 13567, 14567, 23467, 23567, 24567, 34567
\\ \cline{2-2}
{\textbf 5} & 
01247, 01356, 02356, 03456, 03567, 12347, 12457, 12467
\\ \cline{2-2}
&
01256, 01257, 01267, 01346, 01347, 01367, 01467, 01567,
02345, 02347, 02357, 02457, 02567, 03457, 03467, 12345,
12346, 12356, 12456, 12567, 13456, 13467, 23456, 23457
\\ \hline
&
012345, 012346, 012357, 012367, 012456, 013457, 014567, 023467, 024567,
123567, 134567, 234567
\\ \cline{2-2}
\textbf 6&
012567, 013467, 023457, 123456
\\ \cline{2-2}
&
012347, 012356, 012457, 012467, 013456, 013567, 023456, 023567, 034567,
123457, 123467, 124567
\\ \hline
\textbf 7 &
0123456, 0123457, 0123467, 0123567, 0124567, 0134567, 0234567, 1234567
\\ \hline 
\textbf 8 & 01234567 
\\ \hline
\end {tabular}\par
\captionof{table}{The equivalent combination of database for each N. The
equivalent combinations are listed within the same row.}
\label{tab:equivalentdb} 
\bigskip
\end{minipage}

\begin{table}[htb]
  \centering
\begin{subtable}{0.70\linewidth}
  \caption{\small $N$=5, $\zeta+\varphi$=1.7076,$\psi$=0.4510, $w$=01234}
  \label{tab:prob5.1}
\resizebox{\columnwidth}{!}{
\begin{tabular}{|c|c|ccccc|c|ccccc|c|}
\hline
&$\mathcal{A}$&\multicolumn{5}{c|}{$O(\pi)$}&$D(\pi)$&\multicolumn{5}{c|}{$O(u+\varphi)$}&$D(\psi)$\\
\hline
CNOT && 
0 &1 &2 &3 &4 &
&
0 &1 &2 &3 &4 &
\\[-3pt]
\cline{3-7}\cline{9-13}
0&0.7449&0.3463&0.8592&0.5187&0.8197&0.4388&0.2380&0.4388&0.8428&0.4449&0.8375&1.0000&0.9604\\
1&0.8575&1.0000&0.8062&1.0000&0.6723&1.0000&0.3114&1.0000&0.6790&1.0000&0.7084&      &0.9236\\
2&1.0000&      &1.0000&      &1.0000&      &0.5350&      &1.0000&      &1.0000&      &0.9525\\
3&      &      &      &      &      &      &0.7062&      &      &      &      &      &0.9329\\
4&      &&&&&&0.7715&&&&&&0.9845\\
5&&&&&&&0.8948&&&&&&0.9832\\
6&&&&&&&0.9781&&&&&&0.9973\\
7&&&&&&&0.9999&&&&&&0.9990\\
8&&&&&&&1.0000&&&&&&0.9999\\
9&&&&&&&&&&&&&1.0000\\
\hline
\end{tabular}}
\end{subtable}

\begin{subtable}{0.70\linewidth}
  \caption{\small $N$=5, $\zeta+\varphi$=1.7076,$\psi$=0.4510,$w$=01247}
  \label{tab:prob5.2}
\resizebox{\columnwidth}{!}{
\begin{tabular}{|c|c|ccccc|c|ccccc|c|}
\hline
&$\mathcal{A}$&\multicolumn{5}{c|}{$O(\pi)$}&$D(\pi)$&\multicolumn{5}{c|}{$O(u+\varphi)$}&$D(\psi)$\\
\hline
CNOT && 
0 &1 &2 &4 &7 &
&
0 &1 &2 &4 &7 &
\\[-3pt]
\cline{3-7}\cline{9-13}
0&0.6500&1.0000&0.8006&0.5706&0.9271&0.5950&0.2233&0.5950&0.9507&0.6047&0.9005&0.6313&0.9446\\
1&0.6581&      &0.9647&1.0000&0.7872&1.0000&0.1646&1.0000&0.7268&1.0000&0.7588&1.0000&0.3315\\
2&1.0000&      &1.0000&      &1.0000&      &0.3310&      &1.0000&      &1.0000&      &0.9543\\
3&      &      &      &      &      &      &0.4529&      &      &      &      &      &0.8952\\
4&      &&&&&&0.6318&&&&&&0.9766\\
5&      &&&&&&0.7458&&&&&&0.9731\\
6&      &&&&&&0.9251&&&&&&0.9912\\
7&      &&&&&&1.0000&&&&&&0.9979\\
8&&&&&&&&&&&&&0.9999\\
\hline
\end{tabular}}
\end{subtable}

\begin{subtable}{0.70\linewidth}
  \caption{\small $N=5,\zeta+\varphi=1.7076,\psi=0.4510,w=01256$}
  \label{tab:prob5.3}
\resizebox{\columnwidth}{!}{
\begin{tabular}{|c|c|ccccc|c|ccccc|c|}
\hline
&$\mathcal{A}$&\multicolumn{5}{c|}{$O(\pi)$}&$D(\pi)$&\multicolumn{5}{c|}{$O(u+\varphi)$}&$D(\psi)$\\
\hline
CNOT && 
0 &1 &2 &5 &6 &
&
0 &1 &2 &5 &6 &
\\[-3pt]
\cline{3-7}\cline{9-13}
0&0.6542&1.0000&1.0000&0.3625&0.8350&0.3108&0.2221&0.3108&0.8432&0.4082&0.8533&0.2448&0.9604\\
1&0.8575&      &      &1.0000&0.7236&1.0000&0.2538&1.0000&0.6802&1.0000&0.6979&1.0000&0.9279\\
2&0.9397&      &      &      &1.0000&      &0.3660&      &1.0000&      &1.0000&      &0.9717\\
3&1.0000&      &      &      &      &      &0.5617&      &      &      &      &      &0.8967\\
4&0.8184&&&&&&0.7715&&&&&&0.9778\\
5&0.9326&&&&&&0.8026&&&&&&0.9835\\
6&0.9482&&&&&&0.9293&&&&&&0.9979\\
7&1.0000&&&&&&0.9996&&&&&&0.9986\\
8&&&&&&&1.0000&&&&&&0.9998\\
9&&&&&&&&&&&&&0.9999\\
\hline
\end{tabular}
}
\end{subtable}

\begin{subtable}{0.80\linewidth}
  \caption{\small $N=6,\zeta+\varphi=1.8605,\psi=0.8411,w=012345$}
  \label{tab:prob6.1}
\resizebox{\columnwidth}{!}{
\begin{tabular}{|c|c|cccccc|c|cccccc|c|}
\hline
&$\mathcal{A}$&\multicolumn{6}{c|}{$O(\pi)$}&$D(\pi)$&\multicolumn{6}{c|}{$O(u+\varphi)$}&$D(\psi)$\\
\hline
CNOT && 
0 &1 &2 &3 &4 &5 &
&
0 &1 &2 &3 &4 &5 &
\\[-3pt]
\cline{3-8}\cline{10-15}
0&0.8024&0.3570&0.9079&0.3721&0.7899&0.4035&0.7407&0.2733&0.7407&0.3683&0.8720&0.3966&0.8621&0.2421&0.9248\\
1&1.0000&0.3702&0.6940&0.4278&0.7054&1.0000&0.6414&0.1892&0.6414&1.0000&0.6663&1.0000&0.6932&1.0000&0.8756\\
2&      &1.0000&0.6434&1.0000&0.6617&      &1.0000&0.6892&1.0000&      &1.0000&      &1.0000&      &0.9077\\
3&      &      &0.7156&      &0.7156&      &      &0.8268&      &      &      &      &      &      &0.8481\\
4&      &&1.0000&&1.0000&&&1.0000&&&&&&&0.9698\\
5&      &&&&&&&1.0000&&&&&&&0.9806\\
6&      &&&&&&&      &&&&&&&1.0000\\
\hline
\end{tabular}
}
\end{subtable}
\end{table}

\clearpage

\begin{table}
  \centering
  \ContinuedFloat
\begin{subtable}{0.80\linewidth}
  \caption{\small $N=6,\zeta+\varphi=1.8605,\psi=0.8411,w=012347$}
  \label{tab:prob6.2}
\resizebox{\columnwidth}{!}{
\begin{tabular}{|c|c|cccccc|c|cccccc|c|}
\hline
&$\mathcal{A}$&\multicolumn{6}{c|}{$O(\pi)$}&$D(\pi)$&\multicolumn{6}{c|}{$O(u+\varphi)$}&$D(\psi)$\\
\hline
CNOT && 
0 &1 &2 &3 &4 &7 &
&
0 &1 &2 &3 &4 &7 &
\\[-3pt]
\cline{3-8}\cline{10-15}
0&0.7714&0.3917&0.6254&0.3942&0.8433&0.3907&0.8684&0.2694&0.8684&0.4236&0.6345&0.4249&0.8374&0.4950&0.8914\\
1&0.8024&0.5393&0.6601&1.0000&0.8211&1.0000&0.7526&0.3086&0.7526&0.5410&0.6788&1.0000&0.6773&1.0000&0.1893\\
2&1.0000&1.0000&0.5921&      &1.0000&      &1.0000&0.5476&1.0000&1.0000&0.5958&      &1.0000&      &0.9077\\
3&      &      &0.7175&      &      &      &      &0.6324&      &      &0.7258&      &      &      &0.8295\\
4&      &      &1.0000&      &&&&0.7424&&&1.0000&&&&0.9641\\
5&      &&&&&&&0.8726&&&&&&&0.9641\\
6&      &&&&&&&1.0000&&&&&&&0.9824\\
7&&&&&&&&&&&&&&&1.0000\\
\hline
\end{tabular}
}
\end{subtable}

\begin{subtable}{0.80\linewidth}
  \caption{\small $N=6,\zeta+\varphi=1.8605,\psi=0.8411,w=012567$}
  \label{tab:prob6.3}
\resizebox{\columnwidth}{!}{
\begin{tabular}{|c|c|cccccc|c|cccccc|c|}
\hline
&$\mathcal{A}$&\multicolumn{6}{c|}{$O(\pi)$}&$D(\pi)$&\multicolumn{6}{c|}{$O(u+\varphi)$}&$D(\psi)$\\
\hline
CNOT && 
0 &1 &2 &5 &6 &7 &
&
0 &1 &2 &5 &6 &7 &
\\[-3pt]
\cline{3-8}\cline{10-15}
0&0.7500&0.2580&0.7261&0.2841&0.7444&0.4075&0.7368&0.1925&0.7368&0.2322&0.7099&0.3807&0.7493&0.3878&0.9251\\
1&0.7714&1.0000&0.7025&1.0000&0.7219&1.0000&0.7394&0.3731&0.7394&1.0000&0.7227&1.0000&0.7042&1.0000&0.4771\\
2&0.8293&      &1.0000&      &1.0000&      &1.0000&0.2278&1.0000&      &1.0000&      &1.0000&      &0.8899\\
3&1.0000&      &      &      &      &      &      &0.2181&      &      &      &      &      &      &0.9037\\
4&&&&&&&&0.6049&&&&&&&0.9617\\
5&&&&&&&&0.6229&&&&&&&0.9631\\
6&&&&&&&&0.8257&&&&&&&0.9813\\
7&&&&&&&&0.9596&&&&&&&0.9996\\
8&&&&&&&&1.0000&&&&&&&0.9999\\
\hline
\end{tabular}
}
\end{subtable}
\begin{subtable}{0.90\linewidth}
  \caption{\small $N=7,\zeta+\varphi=2.0277,\psi=1.2056,w=0123456$}
\label{tab:prob7}
\resizebox{\columnwidth}{!}{
\begin{tabular}{|c|c|ccccccc|c|ccccccc|c|}
\hline
&$\mathcal{A}$&\multicolumn{7}{c|}{$O(\pi)$}&$D(\pi)$&\multicolumn{7}{c|}{$O(u+\varphi)$}&$D(\psi)$\\
\hline
CNOT && 
0 &1 &2 &3 &4 &5 &6 &
&
0 &1 &2 &3 &4 &5 &6 &
\\[-3pt]
\cline{3-9}\cline{11-17}
0&0.8836&0.5210&0.5794&0.0810&0.4744&0.1625&0.7817&0.4387&0.2131&0.4387&0.6534&0.4318&0.6674&0.4507&0.6005&0.3290&0.8685\\
1&0.8890&0.4903&0.3747&0.4384&0.4728&0.3949&0.4527&1.0000&0.4133&1.0000&0.6564&0.4652&0.4527&1.0000&0.6774&1.0000&0.4798\\
2&0.9149&0.5569&0.4636&1.0000&0.6999&1.0000&0.7098&      &0.5111&      &1.0000&1.0000&0.6985&      &1.0000&      &0.8905\\
3&1.0000&1.0000&0.6879&      &0.7345&      &0.7156&      &0.6613&      &      &      &0.7136&      &      &      &0.8905\\
4&      &      &1.0000&      &1.0000&&0.7825&&0.7016&&&&1.0000&&&&0.9282\\
5&      &&&&&&0.9111&&0.7621&&&&&&&&0.8409\\
6&      &&&&&&1.0000&&0.8239&&&&&&&&0.9754\\
7&&&&&&&&&0.9719&&&&&&&&0.9690\\
8&&&&&&&&&1.0000&&&&&&&&0.9762\\
9&&&&&&&&&      &&&&&&&&1.0000\\
\hline
\end{tabular}
}
\end{subtable}
\begin{subtable}{1.00\linewidth}
  \caption{\small $N=8,\zeta+\varphi=2.2143,\psi=1.5708,w=01234567$}
\label{tab:prob8}
\resizebox{\columnwidth}{!}{
\begin{tabular}{|c|c|cccccccc|c|cccccccc|c|}
\hline
&$\mathcal{A}$&\multicolumn{8}{c|}{$O(\pi)$}&$D(\pi)$&\multicolumn{8}{c|}{$O(u+\varphi)$}&$D(\psi)$\\
\hline
CNOT && 
0 &1 &2 &3 &4 &5 &6 &7 &
&
0 &1 &2 &3 &4 &5 &6 &7 &
\\[-3pt]
\cline{3-10}\cline{12-19}
0&1.0000&0.6193&0.7487&0.3517&0.2817&0.3698&0.7262&0.5016&0.3434&0.0156&0.3434&0.1804&0.6695&0.5933&0.3568&0.3061&0.2336&0.5389&0.8565\\
1&      &0.6296&0.3980&0.1596&0.3178&0.6412&0.3159&0.0864&0.4831&0.7656&0.4831&0.6222&0.5543&0.5685&0.3384&0.1113&0.4132&0.6278&0.6171\\
2&      &0.5042&0.5089&0.3309&0.5023&0.3055&0.5090&0.6350&0.3895&0.5312&0.3895&0.4382&0.3703&0.6321&0.3730&0.6548&0.4884&0.1716&0.8902\\
3&&0.6278&0.7276&0.5495&0.7781&0.5509&0.6162&0.6740&0.7633&0.5312&0.7633&0.5554&0.6200&0.4793&0.6200&0.4275&0.6542&0.7253&0.8902\\
4&&0.5383&0.6945&0.6470&0.7831&0.5704&0.6847&0.4877&0.8043&0.8902&0.8043&0.4531&0.5539&0.6574&0.6994&0.7421&0.6840&0.5769&0.9619\\
5&&0.7407&0.7961&0.7369&0.8875&0.7333&0.8747&0.7372&0.8763&0.8902&0.8763&0.7409&0.8778&0.7286&0.8762&0.7159&0.8760&0.7217&0.9619\\
6&&1.0000&1.0000&1.0000&1.0000&1.0000&1.0000&1.0000&1.0000&1.0000&1.0000&1.0000&1.0000&1.0000&1.0000&1.0000&1.0000&1.0000&1.0000\\
\hline
\end{tabular}
}
\end{subtable}
\caption{The obtained success probabilities $p_s$ of 3-qubit BEQS;  the
corresponding column is approximated with an $l$-size gates network
comprises \{CNOT, $U$\}, where $U$s are 1-qubit unitary gates and $l$ is
the number of CNOT gates.}

\label{tab:prob}
\end{table}

\end{document}